\documentclass[a4paper, oneside, 12pt]{scrartcl}

\usepackage[ngerman,USenglish]{babel}
\usepackage[utf8]{inputenc}
\usepackage{natbib}
\usepackage{amsmath,amssymb,bm}
\usepackage{textcomp}
\usepackage{dsfont}
\usepackage{color}
\usepackage{graphicx}
\usepackage{graphics}
\usepackage{multirow}
\usepackage{lscape}
\usepackage{tabularx}
\usepackage{tabulary}
\usepackage{booktabs}
\usepackage{authblk}
\usepackage{url}
\usepackage{multirow}
\usepackage{rotating}
\usepackage{subcaption}

\usepackage[font={small,sl},labelfont=bf]{caption} 

\usepackage[running]{lineno}
\usepackage[pagebackref]{hyperref}
\hypersetup{%
	colorlinks=true,
	linkcolor=blue, 
	citecolor=blue,
	bookmarksopen=true,
	pdfstartpage={1},
	pdftitle={Title},
	pdfauthor={Arne Nothdurft, Valentin Sarkleti, Tobias Ofner-Graff, Andreas Tockner, Christoph Gollob, Tim Ritter, Ralf Kraßnitzer, Philip Svazek, Martin Kühmaier, Karl Stampfer, Andrew O. Finley},
	pdfkeywords={}
}

\newcommand{\bA}{ {\boldsymbol A} }

\newcommand{\bI}{ {\boldsymbol I} }

\newcommand{\bR}{ {\boldsymbol R} }
\newcommand{\bs}{ {\boldsymbol s} }

\newcommand{\bV}{ {\boldsymbol V} }
\newcommand{\bw}{ {\boldsymbol w} }

\newcommand{\bx}{ {\boldsymbol x} }
\newcommand{\bX}{ {\boldsymbol X} }
\newcommand{\by}{ {\boldsymbol y} }

\newcommand{\bphi}{ {\boldsymbol \phi} }
\newcommand{\bpsi}{ {\boldsymbol \Psi} }

\newcommand{\bbeta}{ {\boldsymbol \beta} }
\newcommand{\btheta}{ {\boldsymbol \theta} }
\newcommand{\bSigma}{ {\boldsymbol \Sigma} }
\newcommand{\bOmega}{ {\boldsymbol \Omega} }
\newcommand{\bepsilon}{ {\boldsymbol \epsilon} }

\newcommand{\bzero}{ {\boldsymbol 0} }

\newcommand{\given}{\,|\,}

\usepackage{letltxmacro}
\LetLtxMacro{\originaleqref}{\eqref}
\renewcommand{\eqref}{Eq.~\originaleqref}

\newcolumntype{L}[1]{>{\raggedright\arraybackslash}p{#1}} 
\newcolumntype{C}[1]{>{\centering\arraybackslash}p{#1}} 
\newcolumntype{R}[1]{>{\raggedleft\arraybackslash}p{#1}} 

\graphicspath{{figures/}}

\title{\Large 
Small area estimation of growing stock timber volume, basal area, mean stem diameter, and stem density for mountain forests in Austria
}

\author[1]{\large Arne Nothdurft\thanks{\href{mailto:arne.nothdurft@boku.ac.at}{arne.nothdurft@boku.ac.at}, Tel: +43-1-47654-91411}}
\author[1]{\large Valentin Sarkleti}
\author[1]{\large Tobias Ofner-Graff}
\author[1]{\large Andreas Tockner}
\author[1]{\large Christoph Gollob}
\author[1]{\large Tim Ritter}
\author[1]{\large Ralf Kra{\ss}nitzer} 
\author[2]{\large Philip Svazek}
\author[2]{\large Martin K{\"u}hmaier}
\author[2]{\large Karl Stampfer} 
\author[3]{\large Andrew O. Finley}

\affil[1]{\normalsize University of Natural Resources and Life Sciences, Vienna (BOKU), Department of Forest and Soil Sciences, Institute of Forest Growth, Austria}

\affil[2]{\normalsize University of Natural Resources and Life Sciences, Vienna (BOKU), Department of Forest and Soil Sciences, Institute of Forest Engineering, Austria}

\affil[3]{\normalsize Michigan State University, Departments of Forestry and Statistics \& Probability, USA}

\date{\large \today}

\begin{document}

\maketitle


\begin{abstract}
\noindent Regression models were evaluated to estimate stand-level growing stock volume (GSV), quadratic mean diameter (QMD), basal area (BA), and stem density (N) in the Brixen im Thale forest district of Austria. Field measurements for GSV, QMD, and BA were collected on 146 inventory plots using a handheld mobile personal laser scanning system. Predictor variables were derived from airborne laser scanning (ALS)–derived normalized digital surface and terrain models. The objective was to generate stand-level estimates and associated uncertainty for GSV, QMD, BA, and N across 824 stands. A unit-level small area estimation framework was used to generate stand-level posterior predictive distributions by aggregating predictions from finer spatial scales. Both univariate and multivariate models, with and without spatially varying intercepts, were considered. Predictive performance was assessed via spatially blocked cross-validation, focusing on bias, accuracy, and precision. Despite exploratory analysis suggesting advantages of complex multivariate spatial models, simpler univariate spatial---and in some cases, non-spatial---models exhibited comparable predictive performance.

\vspace{0.25cm}

\noindent Keywords: Bayesian inference, model-based, forest inventory, airborne laser scanning

\end{abstract}

\section{Introduction}\label{sec:intro}

Sustainable forest management requires precise information on available timber resources. Within forest management planning, this information is used to inform the decision-making process about silvicultural and harvesting activities for individual forest stands. Here, in accordance with \cite{OLIVER1996} and \cite{ASHTON2018}, a stand is defined as the basic unit of management and identified as a spatially continuous extent in which trees are approximately the same age and structure characteristics are homogeneous.

Forest operations that aim to meet desired goals are usually based on criteria associated with stand species composition and structure. For example, mean stand diameter at breast height (DBH) characterizes maturity \citep{SkovsgaardVanclay2007} and is used to differentiate between pre-commercial thinning and final harvesting operations \citep{Kojola2005}. Measures of DBH central tendency, such as mean quadratic DBH (QMD), might also determine technical restrictions for mechanized timber harvesting systems and can serve as a proxy for timber harvesting cost and productivity projections \citep{Eliasson1999, Louis2022}. Basal area (BA) and number of stems per unit area, i.e., stand density, (N) are used to characterize inter-tree competition \citep{Sterba1981, Sterba1985, Sterba1987} and silvicultural prescription harvest removals and residuals. Stand density also influences the efficiency of forest operations. Observed BA is often checked against reference values to estimate whether yield losses are expected due to under- or over-stocking \citep{Assmann1950, Assmann1954}. Finally, the total growing stock timber volume (GSV) is used for timber supply and above-ground carbon estimates \citep{Matala2009}. In summary, stand structure characteristics captured using GSV, QMD, BA, and N are key for defining silvicultural prescriptions, harvesting planning and implementation. \cite{Picchio2020} and \cite{Venanzi2023} show that including more precise data in planning and executing forest operations reduces costs, increases productivity, mitigates soil disturbance, promotes healthier forest ecosystems, and enhances safety and working conditions for forest workers.

Our aim is to generate precise estimates for GSV, QMD, BA, and N at unobserved locations within stands, and summarize these predictions to arrive at stand-level parameter estimates appropriate for supporting silvicultural prescriptions, harvest planning and implementation. Given underlying tree- and stand-level allometric relationships, as well as common growing conditions and disturbance history, we expect these stand parameters to co-vary and exhibit spatial dependence. Chosen estimators should maintain biologically meaningful relationships among outcomes such that they are reflected in subsequent estimates.

Planning and management occur at the stand-level; hence, they need to be informed by stand-level structure parameter estimates. Given cost constraints, installing sufficiently dense forest inventory plot networks to achieve the desired level of estimate precision using traditional design-derived estimators is typically not feasible. As a result, various approaches have been developed to estimate parameters in stands with insufficient numbers of observations. Such approaches commonly couple sparsely sampled forest inventory and remote sensing data within a non-parametric or parametric modeling framework. The efficiency of non-parametric nearest neighbor methods \citep[e.g.,][]{MoeurStage1995, TomppoHalme2004, LeMayTemesgen2005, Maltamo2006}, various interpretations of small area estimators \citep[e.g.,][]{Goerndt2011, Goerndt2012, Breidenbach2012, Mandallaz2013, Steinmann2013, Mandallaz2014, Mauro2016,  Breidenbach2018, Green2019, KatilaHeikkinen2020, Coulston2021, Green2022}, and geostatistical methods \citep[e.g.,][]{Mandallaz2000, Tuominen2003, Zhang2004, Berterretche2005, Buddenbaum2005, Maselli2006, Babcock2018, Finley2024} are explored in many studies and remain an active area of research.

Within our current setting, an advantage of nearest neighbor (NN) methods is that simultaneous imputation of outcome variables/values preserves their covariance; however, a limitation of these methods is there are no exact variance estimators and therefore approximations must be used \citep{McRoberts2007, McRoberts2011, McRoberts2022, Magnussen2013}. Small area estimators that build on stand-level direct estimates (i.e., estimates from design-based methods) are prevalent in the literature; however, they require a sufficient number of observations within each stand to generate direct estimates. Given forest stands are typically only a few hectares in size, there are often too few (or no) observations to achieve the desired level of direct estimate accuracy and precision. Other approaches couple direct with indirect estimators via regression models or NN methods to form composite estimators \citep{Dettmann2022}. While composite estimators might inherit some attractive theoretical properties of the direct estimator used, they too often require prohibitively large samples sizes to deliver reliable estimates, and specification of the model component can be restrictive, especially when extended to multivariate outcome settings. Model-based gostatistical methods offer flexible possibilities to combine spatially independent effects and spatially dependent trends in a coherent model that can be used to predict outcome variables at unobserved locations \citep{Cressie1993, Schabenberger2004}. Further flexibility is achieved when classical geostatistical methods are cast within a Bayesian hierarchical model \citep{Banerjee2014}. Under the Bayesian paradigm, asymptotic distribution assumptions about the parameters are relaxed, and model parameter and predictive inference proceed from posterior and posterior predictive distributions, respectively. 

Several studies assess univariate and multivariate gostatistical Bayesian hierarchical models for forest inventory applications. For example, \cite{Finley2013} and \cite{Babcock2013} compare such models for forest parameter point estimates and mapping. Their work builds on \cite{Finley2008} who detailed  and applied a Bayesian multivariate hierarchical model for predicting forest biomass per unit area of multiple tree species in a 1,053\,ha forest district in the northern USA. Each species' biomass in the outcome vector was represented by a linear model with fixed and spatial random effects. Fixed effects captured the impact of observed environmental variables and spatial effects followed a multivariate spatial Gaussian process (GP) and captured spatial dependence not accounted for by the fixed effects. Covariance among species specific spatial processes was estimated via a linear model of coregionalization \citep{Wackernagel2003, Gelfand2004}. Here too, their model estimated the species specific residual white noise process variance and between species covariance \citep[a model component often refereed to as ``seemingly unrelated regression'' in the forestry literature, see, e.g., ][]{Rose2001}. Their proposed model provides the flexibility and inferential advantages of geostatstical methods and the multivariate GP captures the covariance among outcomes, which is preserved in subsequent predictions. While not explored in \cite{Finley2008} and subsequent applications, the proposed model has ideal qualities for unit-level small area estimation (SAE), which is the focus of our current study.

SAE methods can generally be classified into two groups: unit-level and area-level models. Unit-level models are constructed at the population unit level, which is the minimal unit that can be sampled from a population. Unit-level models characterize the relationship between outcome variables measured on sampled population units with auxiliary data available for all population units. Small area prediction proceeds by summarizing unit-level predictions that fall within the areal extent of interest \citep{rao2015small}. In contrast, area-level models are constructed at the areal unit level, where areal units are pre-defined non-overlapping areas that tessellate the population's extent. Area-level models characterize the relationship between area-specific direct estimates and auxiliary data. Typically, the direct estimates are generated using design-based estimators \citep{rao2015small}. In this way, an area-level model uses available auxiliary information to effectively ``adjust'' the area-specific direct estimates.

In our current study, extreme sparseness of the available forest inventory plots precludes stand-level direct estimates and hence use of area-level models. Rather, we pursue unit-level SAE using Bayesian regression models. Predictive performance is assessed for series of candidate models of increasing complexity. Data used to inform estimates are described in Section~\ref{sec:data} and \ref{sec:variables}, the model definitions, their implementation, and prediction algorithms are given in Section~\ref{sec:model_construction} and \ref{sec:implementation_analysis}. Candidate models and assessment metrics are described in Section~\ref{sec:candidate_models}. Results are given in Section~\ref{sec:results}, followed by the discussion and conclusion in Sections~\ref{sec:discussion} and \ref{sec:conclusion}, respectively.

\section{Materials and methods}\label{sec:material_methods}

\subsection{Study region and data}\label{sec:data}

\begin{figure*}[ht!]
\centering
\includegraphics[width=1.0\textwidth]{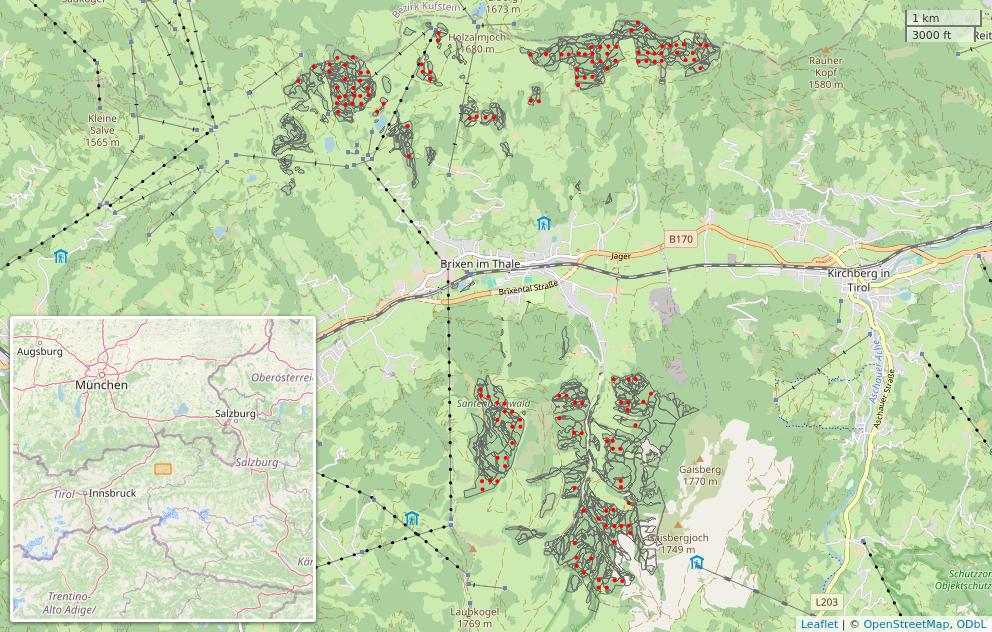}
\caption{Location and extent of the 824 forest stands (grey boundaries) in the forest district Brixen im Thale and locations of the 146 sample points (red dots).}
\label{fig:map_plots_Brixen}
\end{figure*}

The study region is a 548 hectare (ha) municipal forest district in Brixen im Thale located within the administrative district Kitzb\"{u}hel of the Austrian federal state of Tyrol. As shown in Figure~\ref{fig:map_plots_Brixen}, the Brixen forest district is partitioned into 824 forest stands. Summary statistics for the stand areas (ha) are: mean 0.67, median 0.36, standard deviation 1.01, minimum 0.002, and maximum 14.95. 

Forest inventory data were collected on $n$ = 146 sample plots, which were randomly selected from an initially established regular 100\,m$\times$100\,m grid (Figure~\ref{fig:map_plots_Brixen}). Plot data were collected in July 2023 using a handheld mobile personal laser scanner (PLS) GeoSLAM ZEB Horizon (GeoSLAM Ltd., Nottingham, UK). The plot size was not constant. Initial sampling protocol called for a radius of 20 m, but in the very steep terrain and due to the high risk of falling, a smaller plot was measured at some locations. As a result, 59 plots had a 20\,m radius and 87 plots had a 10\,m radius.

Position, DBH, and height for the approximately 7,131 measurement trees having DBH exceeding a lower threshold of 5\,cm were derived from 3D point clouds collected on each plot using fully automated routines detailed in \cite{Gollob2019,Gollob2020a}, \cite{Tockner2022}, and \cite{Ritter2017,Ritter2020}. Stem volume was calculated using a traditional stem-form function \citep{Pollanschuetz1965}. Individual tree detection and tree segmentation from ground-based LiDAR was conducted using the \texttt{treeX} R software package \citep{TocknerTreesX}. 

Laser scanning field work was intentionally conducted using a mobile PLS system rather than a stationary terrestrial laser scanner (TLS). The PLS device is carried through the plot during scanning, enabling trees and their crowns to be captured from multiple angles. This multi-perspective approach results in highly complete point clouds within the crown space, significantly improving the accuracy of both tree height measurements and crown segmentation compared to TLS-derived data. As demonstrated by \cite{Gollob2020a}, the automatic stem detection rate using PLS data reaches 98.5\% for trees with DBH $\geq$ 10,cm and 96\% for trees with DBH $\geq$ 5,cm. Furthermore, automatic height measurements derived from PLS point clouds exhibit no systematic bias and achieve a root mean squared error (RMSE) of only 2.6,m \citep{Tockner2022}.

Based on the maximum tree height and walking path in the area, we estimate the distance between the scanner and the tops of the tree crowns to be about 40\,m. The PLS device has an initial beam size of 12.7\,mm horizontally and 9.5\,mm vertically, with beam divergences of 1.2\,mrad and 3\,mrad, respectively \citep{hypertech_puck_spec_sheet}. These dimensions result in an elliptical footprint of approximately 0.95\,cm\textsuperscript{2} at 0\,m and 45.2\,cm\textsuperscript{2} at 40\,m. While the larger footprint in the upper parts of the tree reduces the ability to resolve small structures, the uncertainty caused by beam divergence ($\pm$6 cm at 40\,m) does not significantly affect the tree height measurement. In fact, the increased footprint makes it more likely that the laser will hit small branches, thereby preserving the completeness of the point cloud, especially in the upper sections of the crowns. However, the accuracy of tree height measurements depends heavily on stand structure. In denser, multilayered forests, occlusion becomes a significant issue as laser pulses are hindered from reaching the treetops. Nonetheless, this problem is far less pronounced for PLS than for TLS, as PLS captures data continuously along the path, unlike TLS, which is limited to a few fixed scan positions \citep{CompeanAguirre2024}. In temperate forests, the accuracy of PLS-derived tree heights is comparable to traditional measurement methods \citep{Jurjevic2020}. However, these results likely do not apply to tropical rainforests.

As described in Section~\ref{sec:intro}, our inferential focus is on GSV, BA, QMD, and N. For subsequent modeling purposes, GSV, BA, and N are expressed on a per hectare basis, e.g., GSV is expressed as m$^3$/ha (i.e., computed as the sum of tree volume on a plot scaled by the plot's tree expansion factor). Summary statistics for the variables are given in Table~\ref{tab:sampleplot_summaries}. 

\begin{table}[ht!]
\caption{Summary statistics from plot measurement data.}
\centering
\label{tab:sampleplot_summaries}
 \begin{tabular}{lrrrrr}
    \toprule
          & mean & median & sd & min & max\\
    \midrule
    GSV (m\textsuperscript{3}/ha) & 510.8 & 461.8 & 257.3 & 109.5 & 1378.5\\
    BA (m\textsuperscript{2}/ha) & 44.8 & 42.0 & 17.9 & 10.5 & 115.5\\
    QMD (cm) & 29.8 & 29.8 & 9.0 & 14.2 & 55.6\\
    N (ha\textsuperscript{-1}) & 790.8 & 700.3 & 489.3 & 71.6 & 2260.0\\
    \bottomrule
\end{tabular}
\end{table}

As noted previously, a key objective in this work is to generate stand-level predictions that preserve the observed correlation among the variables. These correlations, estimated using plot measurement data, are given in Table~\ref{tab:sampleplot_cor}.

\begin{table}[ht!]
\caption{Correlation among variables estimated using plot measurement data.}
\label{tab:sampleplot_cor}
\centering
\begin{tabular}[t]{lcccc}
\toprule
& GSV & QMD & BA & N\\
\midrule
GSV & 1 &  &  & \\
QMD & 0.47 & 1 &  & \\
BA & 0.95 & 0.26 & 1 & \\
N & -0.02 & -0.75 & 0.22 & 1\\
\bottomrule
\end{tabular}
\end{table}

The Austrian federal state of Tyrol provided access to a digital terrain model (DTM) and a digital surface model (DSM) derived from an ALS campaign flown in 2020. DTM and DSM were provided at a 1\,m$\times$1\,m spatial resolution and used to form a normalized digital vegetation height model (DVHM), i.e., by subtracting the DTM from the DSM. To serve as predictor variables in the subsequent model, the mean, standard deviation, and percentiles (denoted later as P$_{x}$ where $x$ is given percentile) were computed using each plot's DVHM values. Similarly, each plot's mean elevation and aspect were computed using DTM values. To help address inconstancy across plot areas, DVHM and DTM values were extracted using a consistent circular area with radius 14.87\,m that was centered on each plot. This radius was root mean square of the plot radii.

\subsection{Outcome and predictor variables}\label{sec:variables}

GSV, BA, and QMD are set as the outcomes in subsequent univariate and multivariate Gaussian models. While we are also interested in N, prior experience has shown that estimating N is challenging. There are a few reasons why our proposed models are not suitable for N. First, because N is a count and better approximated using a Poisson distribution, modeling its distribution as Gaussian might not be appropriate. Second, the ALS-derived DVHM predictor variables hold little information about stem count, especially for understory trees. Hence, the proposed regression models are formulated for GSV, BA, and QMD, and predictions of N are derived from BA's and QMD's predictions as follows. Given DBH, the QMD is defined as
\begin{linenomath*}
 \begin{equation}
\text{QMD} = \sqrt{\frac{1}{\text{N}}\sum_{i=1}^{\text{N}}\text{DBH}_i^2} 
 \label{eq:QMD}
\end{equation}
\end{linenomath*}
and BA is defined as
\begin{linenomath*}
\begin{equation}
\text{BA} = \frac{\pi}{4}\sum_{i=1}^{\text{N}}\left(\frac{{\text{DBH}}_i}{100}\right)^2,
 \label{eq:BA}
\end{equation}
\end{linenomath*}
then rearranging \eqref{eq:QMD} yields
\begin{linenomath*}
\begin{equation}
\text{N}\cdot \text{QMD}^2 = \sum_{i=1}^{\text{N}}\text{DBH}_i^2
 \label{eq:QMD_rear}
\end{equation}
\end{linenomath*}
and replacing \eqref{eq:BA} by \eqref{eq:QMD_rear} gives the desired tree density
\begin{linenomath*}
\begin{equation}
\text{N} = \frac{\text{BA}}{\frac{\pi}{4} \left(\frac{\text{QMD}}{100}\right)^2}
 \label{eq:N}
\end{equation}
\end{linenomath*}
\citep[see, e.g.,][for additional explanation and insights about this relationship]{Robert2000}.

In a preprocessing step, univariate non-spatial regression models were used to select DVHM and DTM predictors. Specifically, forward and backward searches were performed using the \verb|stepAIC| function in the \verb|MASS| R package \citep{Venables2002} to identify a set of predictors that maximized variance explained. Predictors that showed high correlation were removed. The final set of predictors used in the subsequent multivariate model were DVHM's mean, standard deviation (SD), and 95\% height percentile (P$_{95}$), and the DTM's mean elevation and aspect. 

\subsection{Models}\label{sec:model_construction}

\subsubsection{Univariate model}\label{sec:uni_models}

Let $y_q(\bs)$ be one of the outcomes, i.e., GSV, BA, or QMD after appropriate transformation, at generic location $\bs$, where $\bs$ is a vector of easting and northing spatial coordinates. For each $\bs$ within the spatial domain defined by the union of the Brixen forest district stands, we observed a vector of spatially referenced predictors $\bx_q(\bs)$. Here, $\bx_q(\bs)$ is a $1\times p_q$ vector with the first element set to one and subsequent elements set to DVHM and DTM predictor values. Given these data, the posited univariate spatial regression model is 
\begin{linenomath*}
\begin{equation}
y_q(\bs)=\bx_q(\bs)\bbeta_q + w_q(\bs) + \epsilon_q(\bs),
 \label{eq:uni_regression_model}
\end{equation}
\end{linenomath*}
where $\bbeta_q = \left(\beta_1, \beta_2, \ldots, \beta_{p_q}\right)^\top$ is the set of regression parameters corresponding to $\bx_q(\bs)$, $w_q(\bs)$ is a spatial random effect defined below, and $\epsilon_q(\bs)$ is an unstructured residual that follows a zero-mean normal distribution with variance $\tau^2_q$, i.e., $\epsilon_q(\bs)\sim N(0,\tau^2)$.

We assume the spatial random effects follow a zero-mean GP such that for the collection of all $n$ sampling plot locations $\mathcal{S}=\{\bs_1,\bs_2,\ldots,\bs_n\}$ the corresponding $n\times 1$ vector of random effects $\bw_q = \left(w_q(\bs_1), w_q(\bs_2), \ldots, w_q(\bs_n)\right)^\top$ is distributed multivariate normal $MVN\left(\bzero, \bSigma_q(\btheta_q)\right)$, where $\bzero$ is a $n\times 1$ vector of zeros and $\bSigma_q(\btheta_q)$ is a $n\times n$ spatially structured covariance matrix. This covariance matrix is defined as $\sigma^2_q\bR(\phi_q)$ with variance $\sigma^2_q$ and correlation matrix $\bR(\phi_q)$ with decay parameter $\phi_q$. We chose to use an exponential spatial correlation function to define $\bR(\phi_q)$ such that its $(i,j)^{th}$ element equals $\exp(-\phi_q||\bs_i - \bs_j||)$, where $||\bs_i - \bs_j||$ is the Euclidean distance between locations $\bs_i$ and $\bs_j$. The spatial process parameters are collected into $\btheta_q = \left(\sigma^2_q, \phi_q\right)$. Model parameters to be estimated are $\bw_q$ and $\bOmega_q = \left(\bbeta_q, \btheta_q, \tau^2_q\right)$.

To aid interpretation of spatial process parameters, spatial range estimates are reported in terms of an effective spatial range. Because an exponential spatial correlation function is used---which decays rapidly and asymptotically approaches zero with increasing distance---the effective spatial range is defined as the distance at which spatial correlation declines to 0.05. This value provides an interpretable measure of spatial dependence and is expressed in kilometers.

\subsubsection{Multivariate model}\label{sec:multi_models}

We extend the univariate model \eqref{eq:uni_regression_model} to jointly model multiple outcomes. Let $\by(\bs) = \left(y_1(\bs), y_2(\bs), \ldots, y_m(\bs)\right)^\top$ be a vector of values for $m$ outcomes at generic spatial location $\bs$. Here, $\by(\bs)$ is a set of $m=3$ appropriately transformed values of GSV, BA, and QMD. The outcome specific intercept and predictors vector $\bx_q(\bs)$ was defined previously in Section~\ref{sec:uni_models} for $q\in \{1,2,\ldots,m\}$. Given these data, the posited multivariate spatial regression models is
\begin{linenomath*}
\begin{equation}
\by(\bs)=\bX(\bs)\bbeta + \bw(\bs) + \bepsilon(\bs),
 \label{eq:multi_regression_model}
\end{equation}
\end{linenomath*}
where $\bX(\bs)$ is a $m\times p$ block-diagonal design matrix with the $q$\textsuperscript{th} block equal the $1\times p_q$ vector $\bx_q(\bs)$, $\bbeta = \left(\bbeta_1^\top,\ldots,\bbeta_m^\top\right)^\top$ is a $p\times 1$ vector of regression parameters $(p = \sum^m_{q=1}p_q)$ with $\bbeta_q = \left(\beta_1, \ldots, \beta_{p_q}\right)^\top$ being the set of regression parameters corresponding to $\bx_q(\bs)$, $\bw(\bs) = \left(w_1(\bs),\ldots, w_m(\bs)\right)^{\top}$ are spatial random effects, and $\bepsilon(\bs) = \left(\epsilon_1(\bs),\ldots, \epsilon_m(\bs)\right)^{\top}$ are unstructured residuals that follow a zero-mean multivariate normal distribution with a $m\times m$ dispersion matrix $\bpsi$, i.e., $\bepsilon(\bs)\sim MVN(\bzero, \bpsi)$.

We assume the spatial random effects follow a zero-mean multivariate GP such that for the collection of all $n$ sample plot locations $\mathcal{S}$ the corresponding $nm\times 1$ vector of random effects $\bw = \left(\bw(\bs_1)^\top, \bw(\bs_2)^\top, \ldots, \bw(\bs_n)^\top\right)^\top$ is distributed $MVN(\bzero, \bSigma(\btheta))$ where $\bSigma_w(\btheta)$ is the $nm\times nm$ spatially structured covariance matrix. Following \cite{Gelfand2004} and \cite{Finley2008}, $\bSigma(\btheta)$ is a symmetric positive definite covariance matrix with its $(i,j)$\textsuperscript{th} $m\times m$ block equal to $\bA\mathbf{V}(\bs_i,\bs_j; \bphi)\bA^\top$, where $\bA$ is a $m\times m$ lower triangular matrix, $\mathbf{V}(\bs_i,\bs_j; \bphi)$ is a diagonal matrix with diagonal elements $\rho(\bs_i,\bs_j;\phi_q)$, and $\bphi = \{\phi_1,\ldots, \phi_m\}$. The spatial process parameters are collected into $\btheta = \left(\bA, \bphi\right)$. Following from Section~\ref{sec:uni_models}, $\rho(\bs_i,\bs_j;\phi_q)$ is the spatial correlation function $\exp(-\phi_q||\bs_i - \bs_j||)$. When $\bs_i = \bs_j$, i.e., the diagonal blocks in $\bSigma(\btheta)$, $\mathbf{V}(\bs_i,\bs_j; \bphi)$ is the identity matrix and the block equals $\bA\bA^\top$, which is recognized as the within location covariance matrix for the $m$ spatial processes. Model parameters to be estimated are $\bw$ and $\bOmega = \left(\bbeta, \btheta, \bpsi\right)$.

As in the univariate spatial regression, spatial range parameters are reported as effective spatial ranges. Following Eq. (8.2) in \cite{Gelfand2004}, the effective range for each outcome's spatial random effect is derived from the estimated cross-covariance matrix ($\bA\bA^\top$) and the corresponding range parameters ($\bphi$).

\subsection{Parameter estimation and prediction}\label{sec:implementation_analysis}

As outlined in Section~\ref{sec:intro}, the primary inferential objective was to generate stand-level multivariate predictions for each of the 824 forest stands within the Brixen forest district.

To ensure predictions have appropriate positive support and to better satisfy Gaussian model assumptions, outcome variables were log-transformed prior to parameter estimation. Given the data described in Section~\ref{sec:data}, parameters in \eqref{eq:uni_regression_model} and \eqref{eq:multi_regression_model} were estimated using a Bayesian framework. Specifically, the \texttt{spSVC} and \texttt{spMvLM} functions from the \texttt{spBayes} R package \citep{Finley2007, Finley2015} were used for the univariate and multivariate spatial models, respectively. The non-spatial univariate and multivariate models were implemented directly in R. Code and data to fit the proposed models is provided in \cite{NothdurftGit2025}. Inference was based on post burn-in and thinned Markov chain Monte Carlo (MCMC) samples (see Section~\ref{sec:post_inference} for details).

For prediction, the Brixen forest district was divided into 26.36 $\times$ 26.36 m grid cells, matching the area used to compute predictors over the inventory plots. This alignment between observed and prediction units mitigates change-of-support issues \citep[see, e.g.,][]{zhang2024}. Each prediction unit was spatially referenced by its centroid, and DVHM and DTM predictors were computed accordingly.

Following the Bayesian framework, predictive inference is based on posterior predictive distributions (PPDs), which characterize the probabilistic distribution of unobserved outcomes given the observed data and the model structure. PPDs integrate over uncertainty in all model parameters---including regression coefficients, random effects, and variance components---rather than conditioning on fixed point estimates.

Assuming that residual variation for each outcome (i.e., unexplained by predictors) is both correlated across outcomes and spatially structured, two sources of dependence should be accommodated when generating PPDs for GSV, BA, and QMD at the stand level:
\begin{enumerate}
\item Cross-outcome covariance, including both spatial and non-spatial components;
\item Spatial dependence between observed plots and prediction units, as well as among prediction units within each stand.
\end{enumerate}

The multivariate spatial model explicitly accounts for both sources of dependence, whereas the univariate model addresses only the second.
\subsubsection{Univariate model prediction}\label{sec:uni_prediction}

Modeling each outcome separately via the univariate model \eqref{eq:uni_regression_model} precludes explicit estimation of among outcome residual covariance, i.e., we cannot ensure predictions reflect linear relationships seen among observed outcomes. However, outcome specific spatial dependence between observed and prediction units, and among prediction units can be accommodated. 

Within-stand spatial dependence is captured by jointly predicting for all units within a given stand. More specifically, we use $l$ to index stands, i.e., $l\in \{1,2,\ldots,824\}$, and say there are $n_l$ prediction units $\mathcal{S}^\ast_l=\{\bs^\ast_1,\bs^\ast_2,\ldots,\bs^\ast_{n_l}\}$ that comprise the areal extent of stand $\mathcal{B}_l$. Our interest is in sampling from the joint PPD for $\by^\ast_{q,l} = \left(y_q(\bs^\ast_1), y_q(\bs^\ast_2), \ldots, y_q(\bs^\ast_{n_l})\right)^\top$. To do so, we first sample from the random effect's PPD $\bw^\ast_{q,l} = \left(w_q(\bs^\ast_1), w_q(\bs^\ast_2), \ldots, w_q(\bs^\ast_{n_l})\right)^\top$ which is given by 
\begin{equation}
    P\!\left(\bw^\ast_{q,l} \given \text{\emph{Data}}\right) = \int P\!\left(\bw^\ast_{q,l} \given \bw_q, \bOmega_q, \text{\emph{Data}}\right)P\!\left(\bw_q\given \text{\emph{Data}}\right)P\!\left(\bOmega_q\given \text{\emph{Data}}\right)d\bOmega_q d\bw_q,
    \label{eq:uni_w_ppd}
\end{equation}
where \emph{Data} includes observed outcomes and predictors. We need not evaluate \eqref{eq:uni_w_ppd} directly, rather we simulate from $P\!\left(\bw^\ast_{q,l} \given \bw_q, \bOmega_q, \text{\emph{Data}}\right)$ via composition sampling, i.e., given posterior samples $\bw_q^{(t)}$ and $\bOmega_q^{(t)}$ we draw $\bw_{q,l}^{\ast,(t)}$ from $P\!\left(\bw^\ast_{q,l} \given \bw_q^{(t)}, \bOmega_q^{(t)}, \text{\emph{Data}}\right)$ for $t \in \{1, 2, \ldots, T\}$ where $T$ is a sufficiently large number of MCMC samples, see \cite{Finley2007} following Eq. (11) for details. Then, for each $\bw_{q,l}^{\ast,(t)}$ we draw a corresponding $\by_{q,l}^{\ast,(t)}$ from $MVN\left(\bX_{q,l}^\ast\bbeta_q^{(t)} + \bw_{q,l}^{\ast,(t)}, \tau_q^{2,(t)}\bI_l\right)$, where $\bX_{q,l}^\ast$ is the $n_l\times p_q$ design matrix and $\bI_l$ is an $n_l\times n_l$ identity matrix.

\subsubsection{Multivariate model prediction}\label{sec:mult_prediction}

The multivariate model \eqref{eq:multi_regression_model} explicitly estimates covariance among outcome specific spatial random effects and unstructured residuals via its cross-covariance matrix function $\bA\bV(\cdot, \cdot)\bA^\top$ and non-spatial $\bpsi$, respectively. Hence, the model's PPD accounts for both among outcome covariance and, within and among outcome spatial dependence. Following from Section~\ref{sec:uni_prediction}, we sample from the $l^{\text{th}}$ stand's joint PPD $\by^\ast_l = \left(\by(\bs^\ast_1), \by(\bs^\ast_2), \ldots, \by(\bs^\ast_{n_l})\right)^\top$ by first sampling from the random effect's PPD $\bw^\ast_l = \left(\bw(\bs^\ast_1), \bw(\bs^\ast_2), \ldots, \bw(\bs^\ast_{n_l})\right)^\top$ which is given by 
\begin{equation}
    P\!\left(\bw^\ast_l \given \text{\emph{Data}}\right) = \int P\!\left(\bw^\ast_l \given \bw, \bOmega, \text{\emph{Data}}\right)P\!\left(\bw\given \text{\emph{Data}}\right)P\!\left(\bOmega\given \text{\emph{Data}}\right)d\bOmega d\bw.
    \label{eq:mult_w_ppd}
\end{equation}
Again, estimation of \eqref{eq:mult_w_ppd} is done via composition sampling, where we draw $\bw^{\ast,(t)}_l$ from $P\!\left(\bw^\ast_l \given \bw^{(t)}, \bOmega^{(t)}, \text{\emph{Data}}\right)$ for $t \in \{1, 2, \ldots, T\}$. Then, for each $\bw^{\ast,(t)}_l$ we draw a corresponding $\by^{\ast,(t)}_l$ from $MVN\left(\bX^\ast_l\bbeta^{(t)} + \bw^{\ast,(t)}_l, \bI_l\otimes \bpsi^{(t)}\right)$, where $\bX^\ast_l$ is the $n_lm\times p$ design matrix with each $m\times p$ block being the $\bX(\bs^\ast)$ block-diagonal matrix described previously, $\bI_l$ is the previously defined identity matrix, and $\otimes$ is the Kronecker product operator.

\subsubsection{Posterior inference}\label{sec:post_inference}

Parameter posterior inference was conducted using Gibbs sampling and Metropolis-within-Gibbs algorithms for parameters without closed-form full conditional distributions \citep[see, e.g.,][]{RobertsRosenthal2009}. To complete the Bayesian specification of the proposed models, prior distributions were assigned to all model parameters. Priors were chosen to achieve conjugacy where possible, and hyperparameters were selected to be intentionally vague, allowing the data to exert primary influence on the resulting posterior distributions.

For all models, regression coefficients were assigned flat (non-informative) priors. Scalar variance parameters were given inverse-Gamma priors with infinite variance and means centered on the corresponding residual variance estimates from the non-spatial models. Covariance parameters were assigned inverse-Wishart priors, with degrees of freedom and scale hyperparameters selected to be diffuse. Spatial decay parameters were given uniform priors with broad geographic support. Implementation details including the data and code to fit the proposed models are provided in \cite{NothdurftGit2025}.

A total of 500 MCMC batches with 10 iterations each were generated. The first 50\% of the resulting 5,000 samples were discarded as burn-in, and the remaining samples were thinned by a factor of 10, yielding $T = 250$ nearly independent draws.

Because the outcomes were log-transformed prior to fitting \eqref{eq:multi_regression_model}, PPD samples were exponentiated to return them to their original support. This approach is standard in Bayesian analysis and avoids the back-transformation bias that may arise in frequentist settings \citep[see, e.g.,][]{Stow2006}.

Given the $T$ back-transformed PPD samples for GSV, BA, and QMD at each prediction unit, we applied \eqref{eq:N} to generate $T$ corresponding PPD samples for N. For subsequent use, we redefine $\by^{\ast,(t)}_l$ to include the PPD sample for N so that
\begin{equation*}
\by^{\ast,(t)}_l = \left(\by^{\ast,(t)\top}_{\text{GSV},l},\by^{\ast,(t)\top}_{\text{QMD},l},\by^{\ast,(t)\top}_{\text{BA},l},\by^{\ast,(t)\top}_{\text{N},l}\right)^\top
\end{equation*}
with
$\by^{\ast,(t)}_{\text{GSV},l} = \left(y_{\text{GSV}}^{(t)}(\bs^{\ast}_1),\ldots, y_{\text{GSV}}^{(t)}(\bs^{\ast}_{n_l})\right)^\top$, $\by^{\ast,(t)}_{\text{QMD},l} = \left(y_{\text{QMD}}^{(t)}(\bs^{\ast}_1),\ldots, y_{\text{QMD}}^{(t)}(\bs^{\ast}_{n_l})\right)^\top$, $\by^{\ast,(t)}_{\text{BA},l} = \left(y_{\text{BA}}^{(t)}(\bs^{\ast}_1),\ldots, y_{\text{BA}}^{(t)}(\bs^{\ast}_{n_l})\right)^\top$, and $\by^{\ast,(t)}_{\text{N},l} = \left(y_{\text{N}}^{(t)}(\bs^{\ast}_1),\ldots, y_{\text{N}}^{(t)}(\bs^{\ast}_{n_l})\right)^\top$.

Stand-level inference proceeded using the $T$ samples from $\by^\ast_l$'s PPD. Because stand boundaries intersected some prediction units (i.e., creating partial grid cells), samples were weighted by prediction unit area when computing aggregate stand-level PPDs. Specifically, the $t^\text{th}$ PPD sample for the $q^\text{th}$ outcome in the $l^\text{th}$ stand was computed as
\begin{equation}
y_q^{\ast,(t)}(\mathcal{B}_l)=\sum_{i=1}^{n_{l}} a(\bs^{\ast}_i)y_q^{(t)}(\bs^{\ast}_i)/\sum_{i=1}^{n_{l}} a(\bs^{\ast}_i),
    \label{eq:stand_ppd}
\end{equation}
with $a(\bs^{\ast}_i)$ denoting the area of the $i^\text{th}$ cell that falls within the stand boundary.

For each stand, the posterior mean and coefficient of variation (CV, expressed as a percentage) were computed for GSV, BA, QMD, and N and mapped.

\subsection{Candidate models}\label{sec:candidate_models}

The following candidate models are considered in the analysis:

\begin{enumerate}
\item Univariate non-spatial regression, defined by \eqref{eq:uni_regression_model}, with the spatial effect $w_q(\bs)$ omitted.
\item Univariate spatial regression, as specified in \eqref{eq:uni_regression_model}.
\item Multivariate non-spatial regression, defined by \eqref{eq:multi_regression_model}, with the spatial random effects $\bw(\bs)$ omitted.
\item Multivariate spatial regression, as specified in \eqref{eq:multi_regression_model}.
\end{enumerate}

Additional variants of these models are constructed by excluding and including the ALS-derived predictors, referred to as the intercept-only and all-predictors models respectively.

\subsubsection{Evaluating candidate models}\label{sec:methods_kfold}

To assess the fit of candidate models to observed plot measurements, we use the Bayesian $R^2$ introduced by \cite{Gelman2019}. In the spirit of the traditional coefficient of determination used in regression, the Bayesian $R^2$ quantifies the proportion of variance in the outcome explained by the model. It is computed as the variance of the predicted values relative to the total variance (predicted plus residual), averaged over the posterior distribution. We use this metric as a quick and interpretable summary of how well each model fits the observed data. However, it does not reflect the quality of stand-level estimators derived from model predictions.

Each candidate model is treated as a model-based estimator of stand-level means, obtained by aggregating unit-level predictions within each stand, as defined in \eqref{eq:stand_ppd}. However, because observations are only available at the unit level (i.e., individual plots), evaluating the accuracy and reliability of these stand-level estimators poses a challenge.

To assess model properties---specifically bias, accuracy, and precision---we employ cross-validation. A naive approach would use random plot-level cross-validation. While informative for evaluating unit-level predictive performance, this approach risks overestimating the performance of spatial models, as predictions for a held-out plots may be spatially close to observed plots, leading to information leakage and inflated accuracy metrics.

To address this, spatial blocking is used to define cross-validation folds. In this approach, plots are grouped into spatially coherent blocks, and all plots within a block are held out simultaneously. This ensures that predictions are made for spatially distinct areas, aligning more closely with the objective of stand-level inference. By withholding entire spatial blocks, the influence of spatial autocorrelation on model performance metrics is reduced.

Spatial blocks were constructed using the \verb|blockCV| R package \citep{blockCV}, with block sizes set to approximately 250 meters---the average distance between stand centroids and the nearest plot. This scale aligns the folds with the spatial resolution of the target inference. A 20-fold (where folds are blocks) cross-validation was implemented, where each fold comprised a disjoint set of spatial blocks that contained approximately 7 plots. Models were fit using the remaining blocks, and predictions were generated for the held-out plots. These predictions were aggregated to the block level following the procedure in Section~\ref{sec:post_inference}, treating blocks as stand analogues.

\begin{figure}
\begin{center}
\includegraphics[width=1\textwidth]{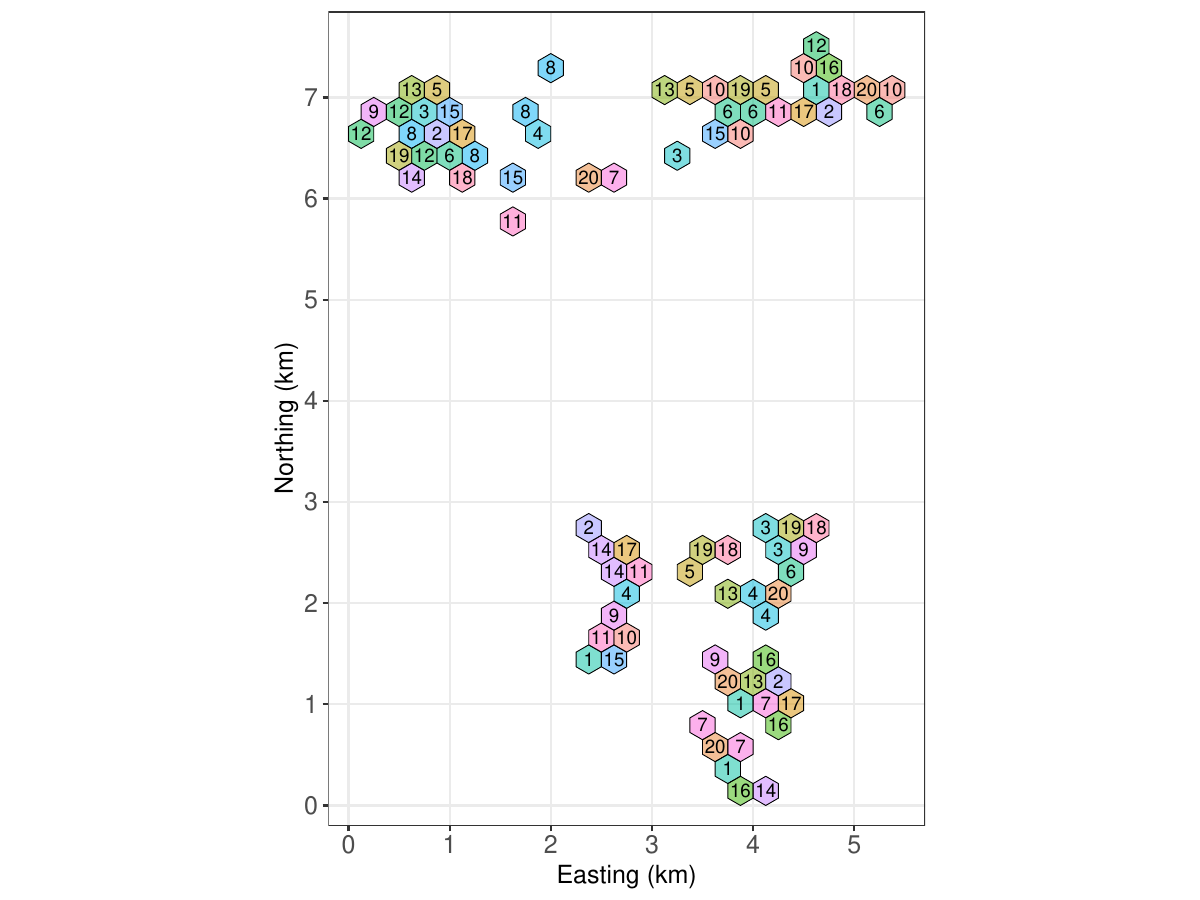}
\caption{Hexagonal blocks used to partition the study area into spatially distinct folds for 20-fold spatial cross-validation. Plots shown in Figure~\ref{fig:map_plots_Brixen} were assigned to their spatially coinciding block. Each block is labeled with its corresponding fold number.}
\label{fig:folds}
\end{center}
\end{figure}

This strategy provides a robust, spatially appropriate evaluation of model-based estimators and facilitates fair comparisons across spatial and non-spatial models. The spatial configuration of the cross-validation blocks is illustrated in Figure~\ref{fig:folds}.

Cross-validation performance was summarized using
\begin{equation}
   \text{bias} = \frac{1}{L}\sum_{l=1}^{L} \left( y_q^{\ast}(\mathcal{B}_l) - y_q(\mathcal{B}_l) \right) 
\end{equation}
and root mean squared prediction error
\begin{equation}
\text{RMSPE} = \sqrt{ \frac{1}{L} \sum_{l=1}^{L} \left( y_q^{\ast}(\mathcal{B}_l) - y_q(\mathcal{B}_l) \right)^2 },
\end{equation}
where $L$ = 20 (the blocks) and, consistent with \eqref{eq:stand_ppd}, the posterior mean for the $l^{\text{th}}$ block is $y_q^{\ast}(\mathcal{B}_l) = \sum_{t = 1}^T\left(\sum_{i=1}^{n_{l}} y_q^{(t)}(\bs^{\ast}_i)/n_l\right)/T$ with $\bs^{\ast}_i$ being the $i^\text{th}$ of the $n_l$ holdout plots that fall with the block. Additional performance metrics included the mean standardized bias and RMSPE (expressed as percentages), empirical coverage of 95\% credible intervals (i.e., the proportion of held-out values captured by their corresponding 95\% PPD intervals), and the average width of these intervals.

Additionally, cross-validation predictions were used to evaluate each estimator's ability to recover the empirical correlations among observed outcomes (Table~\ref{tab:sampleplot_cor}). Given one sample from each outcome's PPD, we computed an among outcome correlation matrix, representing a single draw from the correlation matrix's PPD. Repeating this process yielded a posterior predictive distribution for the correlation matrix. Posterior means were used as point estimates, and 95\% credible intervals were computed to assess whether the empirical correlations in Table~\ref{tab:sampleplot_cor} were captured.

\section{Results}\label{sec:results}

\subsection{Univariate models}

Parameter estimates and model fit summaries for the univariate candidate models described in Section~\ref{sec:uni_models} are presented in Table~\ref{tab:uni_params}. Several regression coefficients identify predictor variables that explain a substantial portion of the variability in the respective outcomes, as indicated by 95\% credible intervals that exclude zero. As noted previously, our primary objective is to develop a predictive model for stand-level parameters; therefore, interpretation of individual regression coefficients is of secondary interest. Moreover, due to potential spatial confounding between covariates and spatial random effects, regression coefficient estimates should be interpreted with caution \citep[see, e.g.,][]{hanks2015,zimmerman2022}.

Comparisons between non-spatial and spatial models reveal consistent evidence of residual spatial dependence across all outcomes. In general, approximately half of the residual variance is attributable to spatial structure in both the intercept-only and all-predictors models. For instance, in the GSV models, the intercept-only non-spatial model yields an estimated residual variance of $\tau^2 \approx 0.28$. In the corresponding spatial model, this variance is partitioned into non-spatial variance $\tau^2 \approx 0.10$ and spatial variance $\sigma^2 \approx 0.18$. The inclusion of ALS-derived predictor variables explains just over half of the residual variance, as reflected by the reduction in $\tau^2$ from the intercept-only to the all-predictors non-spatial model (i.e., from $\sim$0.28 to $\sim$0.1). In the all-predictors spatial model, the residual variance is further divided into $\tau^2 \approx 0.047$ and $\sigma^2 \approx 0.056$. Similar patterns are observed in the QMD and BA models.

The spatial dependence captured by the random effects in both the intercept-only and all-predictors models is characterized by relatively short effective spatial ranges. For example, in the GSV models, the estimated effective spatial range is approximately 0.57~km in the intercept-only model and 0.34~km in the all-predictors model. This indicates limited borrowing of information from observed plots when making predictions beyond these spatial distances.

The reduction in total residual variance (i.e., $\tau^2 + \sigma^2$) and the shortened spatial range observed when moving from the intercept-only to the all-predictors models reflects the spatial explanatory power of ALS-derived predictors. These variables capture underlying spatial patterns in forest structure and, consequently, in forest outcomes.

Across all outcomes, the inclusion of spatial random effects and ALS-derived predictors improves model fit, as evidenced by increasing Bayesian $R^2$ values shown in Table~\ref{tab:uni_params}.

\begingroup
\setlength{\tabcolsep}{1pt}
\renewcommand{\arraystretch}{.75}

\begin{table}[!h]
\centering
\footnotesize
\caption{Posterior summaries for univariate candidate models' parameters. Estimates represent the posterior median with 95\% credible interval (CI) bounds in parentheses. Subscripts on regression coefficients indicate the associated predictor variable. Boldface indicates regression coefficients whose credible intervals exclude zero. Effective range values are in kilometers.}\label{tab:uni_params}
\begin{tabular}{cccccc}
\toprule
\multicolumn{2}{c}{ } & \multicolumn{2}{c}{Intercept only} & \multicolumn{2}{c}{All predictors} \\
\cmidrule(l{3pt}r{3pt}){3-4} \cmidrule(l{3pt}r{3pt}){5-6}
 & Parameter & Non-spatial & Spatial & Non-spatial & Spatial\\
\midrule
 & $\beta_1$ & \textbf{6.1 (6.0, 6.2)} & \textbf{6.1 (6.0, 6.3)} & \textbf{3.8 (3.2, 4.5)} & \textbf{3.9 (3.1, 4.7)}\\
 & $\beta_{\text{Mean}}$ & - & - & \textbf{0.080 (0.058, 0.10)} & \textbf{0.080 (0.058, 0.10)}\\
 & $\beta_{\text{SD}}$ & - & - & 0.015 (-0.032, 0.063) & 0.022 (-0.026, 0.068)\\
 & $\beta_{\text{P}_{95}}$ & - & - & -0.0015 (-0.032, 0.029) & -0.0052 (-0.035, 0.025)\\
 & $\beta_{\text{Elev.}}$ & - & - & \textbf{0.064 (0.022, 0.11)} & \textbf{0.064 (0.010, 0.11)}\\
 & $\beta_{\text{Asp.}}$ & - & - & -0.024 (-0.12, 0.081) & -0.031 (-0.14, 0.074)\\
 & $\tau^2$ & 0.28 (0.22, 0.36) & 0.10 (0.025, 0.20) & 0.10 (0.080, 0.13) & 0.047 (0.019, 0.084)\\
 & $\sigma^2$ & - & 0.18 (0.074, 0.31) & - & 0.056 (0.024, 0.098)\\
 & Eff. range & - & 0.57 (0.26, 1.7) & - & 0.34 (0.12, 1.1)\\
  \cmidrule{2-6}
\multirow{-10}{*}{\centering\arraybackslash \rotatebox{90}{GSV}} & $R^2$ & 0.0070 (0.0055, 0.0089) & 0.61 (0.28, 0.90) & 0.65 (0.57, 0.71) & 0.83 (0.71, 0.93)\\
\cmidrule{1-6}
 & $\beta_1$ & \textbf{3.3 (3.3, 3.4)} & \textbf{3.4 (3.3, 3.5)} & \textbf{1.6 (1.2, 2.0)} & \textbf{1.6 (0.98, 2.2)}\\
 & $\beta_{\text{Mean}}$ & - & - & \textbf{0.033 (0.020, 0.045)} & \textbf{0.036 (0.022, 0.050)}\\
 & $\beta_{\text{SD}}$ & - & - & \textbf{0.088 (0.062, 0.11)} & \textbf{0.095 (0.068, 0.12)}\\
 & $\beta_{\text{P}_{95}}$ & - & - & -0.012 (-0.029, 0.0062) & -0.016 (-0.035, 0.0024)\\
 & $\beta_{\text{Elev.}}$ & - & - & \textbf{0.056 (0.031, 0.079)} & \textbf{0.060 (0.022, 0.10)}\\
 & $\beta_{\text{Asp.}}$ & - & - & \textbf{0.065 (0.0079, 0.12)} & 0.042 (-0.028, 0.11)\\
 & $\tau^2$ & 0.095 (0.077, 0.12) & 0.054 (0.023, 0.083) & 0.034 (0.028, 0.044) & 0.021 (0.013, 0.030)\\
 & $\sigma^2$ & - & 0.047 (0.022, 0.089) & - & 0.021 (0.012, 0.039)\\
 & Eff. range & - & 0.79 (0.22, 3.1) & - & 1.1 (0.19, 3.7)\\
 \cmidrule{2-6}
\multirow{-10}{*}{\centering\arraybackslash \rotatebox{90}{QMD}} & $R^2$ & 0.0072 (0.0057, 0.0090) & 0.43 (0.22, 0.72) & 0.65 (0.56, 0.72) & 0.62 (0.48, 0.80)\\
\cmidrule{1-6}
 & $\beta_1$ & \textbf{3.7 (3.7, 3.8)} & \textbf{3.7 (3.6, 3.9)} & \textbf{2.2 (1.5, 2.9)} & \textbf{2.3 (1.4, 3.0)}\\
 & $\beta_{\text{Mean}}$ & - & - & \textbf{0.053 (0.031, 0.075)} & \textbf{0.053 (0.030, 0.075)}\\
 & $\beta_{\text{SD}}$ & - & - & -0.033 (-0.081, 0.012) & -0.025 (-0.069, 0.020)\\
 & $\beta_{\text{P}_{95}}$ & - & - & 0.0037 (-0.027, 0.035) & 0.00068 (-0.029, 0.032)\\
 & $\beta_{\text{Elev.}}$ & - & - & \textbf{0.062 (0.020, 0.10)} & \textbf{0.060 (0.0090, 0.11)}\\
 & $\beta_{\text{Asp.}}$ & - & - & -0.014 (-0.11, 0.088) & -0.027 (-0.12, 0.078)\\
 & $\tau^2$ & 0.18 (0.14, 0.23) & 0.069 (0.022, 0.13) & 0.096 (0.077, 0.12) & 0.045 (0.018, 0.083)\\
 & $\sigma^2$ & - & 0.11 (0.048, 0.19) & - & 0.055 (0.023, 0.094)\\
 & Eff. range & - & 0.57 (0.26, 1.7) & - & 0.34 (0.13, 0.95)\\
  \cmidrule{2-6}
\multirow{-10}{*}{\centering\arraybackslash \rotatebox{90}{BA}} & $R^2$ & 0.0067 (0.0052, 0.0083) & 0.59 (0.28, 0.88) & 0.49 (0.37, 0.58) & 0.74 (0.60, 0.88)\\
\bottomrule
\end{tabular}
\end{table}
\endgroup

\begin{figure}[!h]
\begin{center}
\includegraphics[width=12cm]{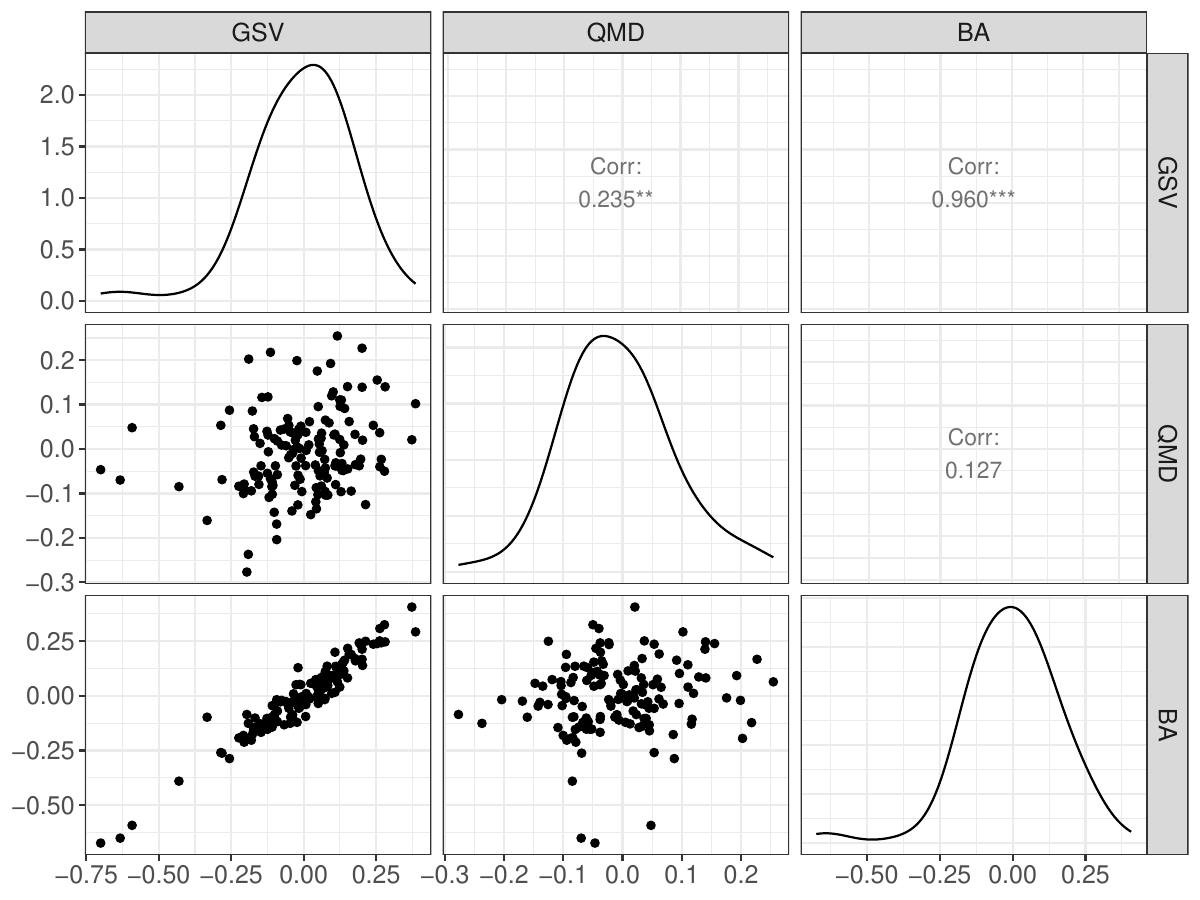}
\caption{Summaries of residuals from univariate spatial all-predictors candidate models. Diagonal panels show the distribution of each model’s spatial random effect; lower triangle panels display pairwise scatter plots between models; and upper triangle panels report the corresponding Pearson correlation coefficients. Asterisks next to the correlation estimates indicate frequentist statistical significance: \textit{**} $p < 0.01$, and \textit{***} $p < 0.001$.}
\label{fig:uni_random_effects_scatter}
\end{center}
\end{figure}

Figure~\ref{fig:uni_random_effects_scatter} shows the distribution of spatial random effects from the all-predictors models, along with scatter plots and pairwise empirical correlations among them. A strong correlation is evident between GSV and BA, while correlations between GSV and QMD, and between QMD and BA, are weak. Similar correlation patterns are observed in the models' residuals (observed minus fitted values), as shown in Figure~\ref{fig:uni_resid_scatter}. These few strong correlations motivate further exploration using the multivariate regression models introduced in Section~\ref{sec:multi_models}, which explicitly estimate such linear associations through the spatial cross-covariance matrix and the non-spatial residual covariance matrix.

\subsection{Multivariate models}

Regression coefficient estimates for the multivariate intercept-only and all-predictors models described in Section~\ref{sec:multi_models} are nearly identical to their univariate counterparts shown in Table~\ref{tab:uni_params}, and are provided in Table~\ref{tab:multi_beta_params}. As before, interpretation of these parameters is of secondary interest---we are primarily interested in predictive performance.

Process parameter estimates and model fit summaries for the multivariate candidate models are presented in Table~\ref{tab:multi_process_params}. Off-diagonal elements of the spatial cross-covariance matrix $\bA\bA^\top$ and the residual covariance matrix $\bpsi$ with 95\% credible intervals that exclude zero are highlighted in bold. As expected, these strong covariances correspond to the same relationships identified through the univariate exploratory scatter plots shown in Figures~\ref{fig:uni_random_effects_scatter} and \ref{fig:uni_resid_scatter}.

The $R^2$ estimates reported at the bottom of Table~\ref{tab:multi_process_params} are generally consistent with their univariate counterparts in Table~\ref{tab:uni_params}, with the exception that the multivariate spatial models yield notably higher $R^2$ values. This improved fit to the observed data is not unexpected, as the multivariate models account for residual covariance among outcomes and may therefore better approximate the true data-generating process. Whether this improvement in fit translates to better predictive performance is evaluated through the cross-validation results presented in Section~\ref{sec:results_cross_validation}.

\begingroup
\setlength{\tabcolsep}{1pt}
\renewcommand{\arraystretch}{.75}
\begin{sidewaystable}[!h]
\centering
\footnotesize
\caption{Posterior summaries for multivariate candidate models' process parameters and model fit $R^2$. Estimates represent the posterior median with 95\% credible interval (CI) bounds in parentheses. Where appropriate, boldface indicates whose credible intervals exclude zero. Effective range values are in kilometers.}\label{tab:multi_process_params}
\begin{tabular}{ccccc}
\toprule
\multicolumn{1}{c}{ } & \multicolumn{2}{c}{Intercept only} & \multicolumn{2}{c}{All predictors} \\
\cmidrule(l{3pt}r{3pt}){2-3} \cmidrule(l{3pt}r{3pt}){4-5}
Parameter & Non-spatial & Spatial & Non-spatial & Spatial\\
\midrule
$\mathbf{\Psi}_{\mathrm{GSV}}$ & 0.28 (0.22, 0.35) & 0.065 (0.014, 0.17) & 0.10 (0.082, 0.13) & 0.030 (0.0082, 0.083)\\
$\mathbf{\Psi}_{\mathrm{QMD,GSV}}$ & \textbf{0.080 (0.053, 0.11)} & 0.032 (-0.0019, 0.069) & 0.0087 (-0.0014, 0.020) & -0.00031 (-0.013, 0.011)\\
$\mathbf{\Psi}_{\mathrm{BA,GSV}}$ & \textbf{0.21 (0.17, 0.27)} & \textbf{0.038 (0.0021, 0.12)} & \textbf{0.094 (0.075, 0.12)} & \textbf{0.029 (0.0045, 0.079)}\\
$\mathbf{\Psi}_{\mathrm{QMD}}$ & 0.097 (0.077, 0.12) & 0.056 (0.019, 0.088) & 0.034 (0.027, 0.043) & 0.021 (0.013, 0.030)\\
$\mathbf{\Psi}_{\mathrm{BA,QMD}}$ & \textbf{0.031 (0.011, 0.054)} & 0.0037 (-0.022, 0.034) & -0.00094 (-0.011, 0.0088) & -0.0084 (-0.021, 0.0029)\\

$\mathbf{\Psi}_{\mathrm{BA}}$ & 0.18 (0.14, 0.23) & 0.034 (0.0074, 0.10) & 0.098 (0.078, 0.12) & 0.036 (0.0082, 0.084)\\
$\bA\bA^\top_{\mathrm{GSV}}$ & - & 0.22 (0.12, 0.43) & - & 0.072 (0.023, 0.11)\\
$\bA\bA^\top_{\mathrm{QMD,GSV}}$ & - & 0.048 (-0.0022, 0.10) & - & 0.0080 (-0.0065, 0.024)\\
$\bA\bA^\top_{\mathrm{BA,GSV}}$ & - & \textbf{0.17 (0.092, 0.36)} & - & \textbf{0.065 (0.018, 0.10)}\\
$\bA\bA^\top_{\mathrm{QMD}}$ & - & 0.047 (0.016, 0.088) & - & 0.017 (0.010, 0.032)\\
$\bA\bA^\top_{\mathrm{BA,QMD}}$ & - & 0.025 (-0.013, 0.064) & - & 0.0041 (-0.013, 0.019)\\
$\bA\bA^\top_{\mathrm{BA}}$ & - & 0.15 (0.084, 0.32) & - & 0.069 (0.021, 0.11)\\
GSV Eff. range & - & 0.46 (0.22, 1.3) & - & 0.30 (0.12, 0.99)\\
QMD Eff. range & - & 0.93 (0.27, 3.1) & - & 1.2 (0.21, 3.3)\\
BA Eff. range & - & 0.75 (0.42, 1.6) & - & 1.1 (0.41, 2.4)\\
\cmidrule{1-5}
GSV $R^2$ & 0.0068 (0.0053, 0.0084) & 0.77 (0.44, 0.95) & 0.65 (0.57, 0.71) & 0.89 (0.72, 0.97)\\
QMD $R^2$ & 0.0067 (0.0053, 0.0084) & 0.42 (0.17, 0.81) & 0.66 (0.57, 0.72) & 0.79 (0.70, 0.87)\\
BA $R^2$ & 0.0068 (0.0055, 0.0086) & 0.81 (0.45, 0.96) & 0.48 (0.37, 0.58) & 0.81 (0.56, 0.95)\\
\bottomrule
\end{tabular}
\end{sidewaystable}
\endgroup

\subsection{Cross-validation}\label{sec:results_cross_validation}

Results from the spatially blocked 20-fold cross-validation, described in Section~\ref{sec:methods_kfold}, are presented in Table~\ref{tab:kfold_metrics}. These performance metrics aim to assess how well each candidate model performs as an estimator of stand-level means. As described in \eqref{eq:stand_ppd}, these stand-level estimates are obtained by aggregating unit-level predictions within each stand.

Where applicable, metric scores are highlighted in bold to indicate the best-performing model. Overall, the results are mixed: no single model within the intercept-only or all-predictors consistently outperforms the others across all outcomes or performance metrics.

Beginning with bias and mean standardized percentage bias. Percentage bias across the three modeled outcomes (GSV, QMD, and BA) is generally small. The absolute values range from 0.14\% for the intercept-only multivariate non-spatial QMD model to 2.0\% for the all-predictors univariate spatial GSV model. There is some indication of a trend toward negative bias in the all-predictors models. In contrast, for the derived outcome $N$, percentage bias is relatively large, ranging in absolute value from 3.8\% to 10\%. However, across all outcomes, the differences in bias scores between models are relatively minor, and no model consistently performs best in terms of bias.

RMSPE and mean standardized percent RMSPE scores favor models that incorporate more information, particularly those with spatial random effects and predictor variables. Specifically, for GSV and QMD, the addition of the ALS-derived predictors reduce RMSPE by half that of the corresponding intercept-only model. For BA, the reduction in RMSPE with addition of predictors is about two-thirds that of the corresponding intercept-only model. The largest predictive gains come from the inclusion of ALS-derived predictors, while the added benefit from spatial random effects is modest. Furthermore, there is no consistent evidence of improved predictive performance when moving from univariate to multivariate spatial models.

The empirical 95\% credible interval (CI) coverage rates do not raise immediate concerns, though they should be interpreted with caution for two reasons. First, in a well-calibrated Bayesian model, we expect 95\% posterior predictive credible intervals to contain new observations approximately 95\% of the time. However, this nominal coverage may not directly extend to aggregated predictions, such as the stand-level means considered here. Second, the reported coverage rates are estimated from a relatively small holdout set (20 folds), introducing uncertainty into the empirical rates. That said, coverage rates substantially below $\sim$90\% would raise concerns about the model’s ability to quantify predictive uncertainty.

Given acceptable bias, accuracy, and coverage rates, a preferable model is one that provides narrower credible intervals, indicating greater predictive precision. The 95\% CI width scores suggest that the all-predictors models yield substantially more precise predictions than the intercept-only models.

\begingroup
\setlength{\tabcolsep}{3.5pt}
\renewcommand{\arraystretch}{.75}
\begin{sidewaystable}[!h]
\footnotesize
\centering
\caption{Block-level cross-validation summary for candidate models. Where applicable, the best-performing value for each metric is shown in bold. In cases where multiple models yield similar performance, more than one value may be bolded.}\label{tab:kfold_metrics}
\begin{tabular}{cccccccccc}
\toprule
\multicolumn{2}{c}{ } & \multicolumn{4}{c}{Intercept only} & \multicolumn{4}{c}{All predictors} \\
\cmidrule(l{3pt}r{3pt}){3-6} \cmidrule(l{3pt}r{3pt}){7-10}
\multicolumn{2}{c}{ } & \multicolumn{2}{c}{Univariate} & \multicolumn{2}{c}{Multivariate} & \multicolumn{2}{c}{Univariate} & \multicolumn{2}{c}{Multivariate} \\
\cmidrule(l{3pt}r{3pt}){3-4} \cmidrule(l{3pt}r{3pt}){5-6} \cmidrule(l{3pt}r{3pt}){7-8} \cmidrule(l{3pt}r{3pt}){9-10}
 & Metric & Non-spatial & Spatial & Non-spatial & Spatial & Non-spatial & Spatial & Non-spatial & Spatial\\
\midrule
 & Bias & -2.5 & -2.0 & -3.4 & \textbf{0.44} & -9.5 & -10. & -8.6 & -8.0\\

 & Bias (\%) & -0.48 & -0.39 & -0.66 & \textbf{0.086} & -1.8 & -2.0 & -1.7 & -1.6\\

 & RMSPE & 96. & 88. & 96. & 88. & 53. & 48. & 52. & \textbf{47.}\\

 & RMSPE (\%) & 19. & 17. & 19. & 17. & 10. & 9.3 & 10. & \textbf{9.0}\\

 & 95\% Cover & 100. & 100. & 95. & 100. & 95. & 100. & 95. & 100.\\

\multirow{-6}{*}{\centering\arraybackslash \rotatebox{90}{GSV}} & 95\% CI Range & 440. & 420. & 440. & 410. & 280. & 290. & 290. & \textbf{280.}\\
\cmidrule{1-10}
 & Bias & 0.069 & 0.067 & \textbf{0.042} & 0.12 & -0.070 & -0.21 & -0.085 & -0.16\\

 & Bias (\%) & 0.23 & 0.22 & \textbf{0.14} & 0.39 & -0.23 & -0.70 & -0.29 & -0.52\\

 & RMSPE & 2.6 & 2.7 & 2.6 & 2.6 & \textbf{1.4} & \textbf{1.4} & 1.4 & 1.4\\

 & RMSPE (\%) & 8.7 & 9.0 & 8.8 & 8.7 & \textbf{4.6} & \textbf{4.6} & 4.7 & 4.7\\

 & 95\% Cover & 95. & 100. & 100. & 100. & 95. & 100. & 95. & 100.\\

\multirow{-6}{*}{\centering\arraybackslash \rotatebox{90}{QMD}} & 95\% CI Range & 14. & 14. & 14. & 13. & 8.7 & 9.0 & 8.8 & \textbf{8.6}\\
\cmidrule{1-10}
 & Bias & -0.19 & \textbf{-0.12} & -0.32 & -0.075 & -0.57 & -0.75 & -0.57 & -0.60\\

 & Bias (\%) & -0.42 & \textbf{-0.27} & -0.70 & -0.17 & -1.3 & -1.7 & -1.3 & -1.3\\

 & RMSPE & 6.5 & 5.8 & 6.5 & 5.8 & 4.7 & \textbf{4.4} & 4.8 & 4.5\\

 & RMSPE (\%) & 14. & 13. & 15. & 13. & 10. & \textbf{9.8} & 11. & 9.9\\

 & 95\% Cover & 95. & 100. & 95. & 100. & 90. & 100. & 90. & 100.\\

\multirow{-6}{*}{\centering\arraybackslash \rotatebox{90}{BA}} & 95\% CI Range & 30. & 29. & 30. & 28. & 23. & 24. & 23. & \textbf{22.}\\
\cmidrule{1-10}
 & Bias & -79. & -79. & \textbf{-30.} & -38. & \textbf{-30.} & -32. & -31. & -37.\\

 & Bias (\%) & -10. & -10. & \textbf{-3.8} & -4.8 & \textbf{-3.8} & -4.0 & -3.9 & -4.7\\

 & RMSPE & 140. & 160. & 130. & 130. & \textbf{100.} & \textbf{100.} & \textbf{100.} & 110.\\

 & RMSPE (\%) & 18. & 20. & 16. & 17. & \textbf{13.} & \textbf{13.} & \textbf{13.} & 14.\\

 & 95\% Cover & 100. & 100. & 100. & 100. & 100. & 100. & 100. & 100.\\

\multirow{-6}{*}{\centering\arraybackslash \rotatebox{90}{N}} & 95\% CI Range & 1100. & 1100. & 900. & 880. & \textbf{680.} & 700. & 700. & 720.\\
\bottomrule
\end{tabular}
\end{sidewaystable}
\endgroup

While not our primary inferential objective, it may be instructive to consider the spatially blocked cross-validation metrics computed for unit-level predictions, which are presented in Table~\ref{tab:kfold_unit_metrics}. These scores are useful for evaluating models intended for point prediction within stands and to assess the appropriateness of the posited models for these unit-support data. The unit-level metric scores exhibit patterns consistent with those observed in the aggregated block-level results shown in Table~\ref{tab:kfold_metrics}. In particular, bias is generally low across models; the inclusion of predictor variables improves both accuracy and precision; and the addition of spatial random effects---whether univariate or multivariate---yields only marginal, if any, performance gains. At the unit level, where predictions directly align with the model structure, we would expect approximately 95\% nominal coverage from the 95\% credible intervals, which is indeed observed across all models. 

These unit-level cross-validation metrics also provide insight into the appropriateness of the posited models. In particular, when defining the spatial processes, we assumed stationary GPs. If such assumptions about the error structure were grossly inappropriate, we would expect to see systematic deviations from nominal predictive coverage. However, the observed coverage rates remain consistently close to 95\% across all models, suggesting the spatial variance components are well estimated and the stationarity assumption is not leading to biased or overconfident uncertainty estimates. While this does not preclude the possibility of localized departures from stationarity, these results suggest the stationary GP assumption is sufficiently flexible and performs well for the purposes of unit-level prediction and uncertainty quantification in this context.

Formal tests for non-stationarity do exist \citep[see, e.g.,][]{Fuentes2005}, but they typically require large spatial domains and dense sampling to reliably detect departures from stationarity---conditions that are not met in the present setting. Even if moderate departures from stationarity were present, it is unclear whether introducing a more complex error structure would yield meaningful gains in predictive performance.

\begin{table}[htbp]
\centering
\caption{Correlation estimates between outcomes based on the cross-validation for all-predictors models.}\label{tab:cor_all}
\begin{subtable}[t]{0.45\textwidth}
  \centering
  \caption{Univariate non-spatial}
  \label{tab:cor_all_a}
  \begin{tabular}{lllll}
  \toprule
  & \text{GSV} & \text{QMD} & \text{BA} & \text{N} \\
  \text{GSV} & 1.0 &  &  & \\
  \text{QMD} & 0.36$^\ast$ & 1.0 &  & \\
  \text{BA} & 0.53 & 0.20$^\ast$ & 1.0 & \\
  \text{N} & 0.0058$^\ast$ & -0.63 & 0.43 & 1.0\\
  \bottomrule
  \end{tabular}
\end{subtable}
\hfill
\begin{subtable}[t]{0.45\textwidth}
  \centering
  \caption{Univariate spatial}
  \label{tab:cor_all_b}
  \begin{tabular}{lllll}
    \toprule
  & \text{GSV} & \text{QMD} & \text{BA} & \text{N} \\
\text{GSV} & 1.0 &  &  & \\
\text{QMD} & 0.36$^\ast$ & 1.0 &  & \\
\text{BA} & 0.52 & 0.19$^\ast$ & 1.0 & \\
\text{N} & -0.0083$^\ast$ & -0.64 & 0.41 & 1.0\\
  \bottomrule
  \end{tabular}
\end{subtable}
\vskip1em
\begin{subtable}[t]{0.45\textwidth}
  \centering
  \caption{Multivariate non-spatial}
  \label{tab:cor_all_c}
  \begin{tabular}{lllll}
  \toprule
  & \text{GSV} & \text{QMD} & \text{BA} & \text{N} \\
  \text{GSV} & 1.0 &  &  & \\
  \text{QMD} & 0.41$^\ast$ & 1.0 &  & \\
  \text{BA} & 0.94$^\ast$ & 0.19$^\ast$ & 1.0 & \\
  \text{N} & 0.20 & -0.63 & 0.43 & 1.0\\
  \bottomrule
  \end{tabular}
\end{subtable}
\hfill
\begin{subtable}[t]{0.45\textwidth}
  \centering
  \caption{Multivariate spatial}
  \label{tab:cor_all_d}
  \begin{tabular}{lllll}
  \toprule
  & \text{GSV} & \text{QMD} & \text{BA} & \text{N} \\
  \text{GSV} & 1.0 &  &  & \\
  \text{QMD} & 0.40$^\ast$ & 1.0 &  & \\
  \text{BA} & 0.93$^\ast$ & 0.16$^\ast$ & 1.0 & \\
  \text{N} & 0.18$^\ast$ & -0.63 & 0.45 & 1.0\\
  \bottomrule
  \end{tabular}
\end{subtable}
\caption*{\small Values are posterior means. An asterisk (\,$^\ast$) indicates that the 95\% credible interval includes the corresponding empirical correlation given in Table~\ref{tab:sampleplot_cor}.}
\end{table}

A key objective of this study was to assess the ability of candidate models to preserve the observed correlations among outcomes, which are assumed to be biologically meaningful. These empirical correlations, derived from the plot data, are shown in Table~\ref{tab:sampleplot_cor}. Table~\ref{tab:cor_all} presents the corresponding correlations estimated during cross-validation using the all-predictors models. For brevity, results from the intercept-only models are reported separately in Table~\ref{tab:cor_int}.

As shown in Tables~\ref{tab:cor_all_a} and \ref{tab:cor_all_b}, the univariate models incorporating predictor variables were able to accurately estimate three of the six pairwise outcome correlations---where ``accurate'' is defined as the observed correlation falling within the estimated 95\% credible interval. Notably, however, these univariate models failed to capture the strong correlation between GSV and BA. The multivariate non-spatial model (Table~\ref{tab:cor_all_c}) similarly captured three of the six correlations well, including the strong relationship between GSV and BA, but failed to capture the correlations involving N. The multivariate spatial model (Table~\ref{tab:cor_all_d}) performed best, accurately estimating four of the six correlations, including the strong GSV and BA relationship. 

In contrast, the intercept-only univariate models (Tables~\ref{tab:cor_int_a} and \ref{tab:cor_int_b}) failed to capture any of the empirical correlations, as they lack both predictor information and any structure to induce correlation among residuals or random effects. Introducing a residual correlation structure via the $\Psi$ matrix in the intercept-only multivariate non-spatial model (Table~\ref{tab:cor_int_c}), and combining this with spatial random effects in the intercept-only multivariate spatial model (Table~\ref{tab:cor_int_d}), substantially improved the ability to estimate observed outcome correlations. However, it is important to emphasize that while the inclusion of $\Psi$ does improve the modeled correlations, the overall model performance---judged by $R^2$ and cross-validation metrics---remains poor.

\subsection{Spatial prediction}\label{sec:spatial_prediction}

Given the results presented in Section~\ref{sec:results_cross_validation}, we expect all candidate models to yield similar levels of predictive performance, provided they are informed by the predictor variables. For illustrative purposes, we present maps of stand-level predictions generated using the all-predictors multivariate spatial model. Maps produced by the other all-predictors models are qualitatively similar.

\begin{landscape}
\begin{figure*}
\begin{minipage}[c]{0.35\textheight}\centering
\includegraphics[width=1\textwidth]{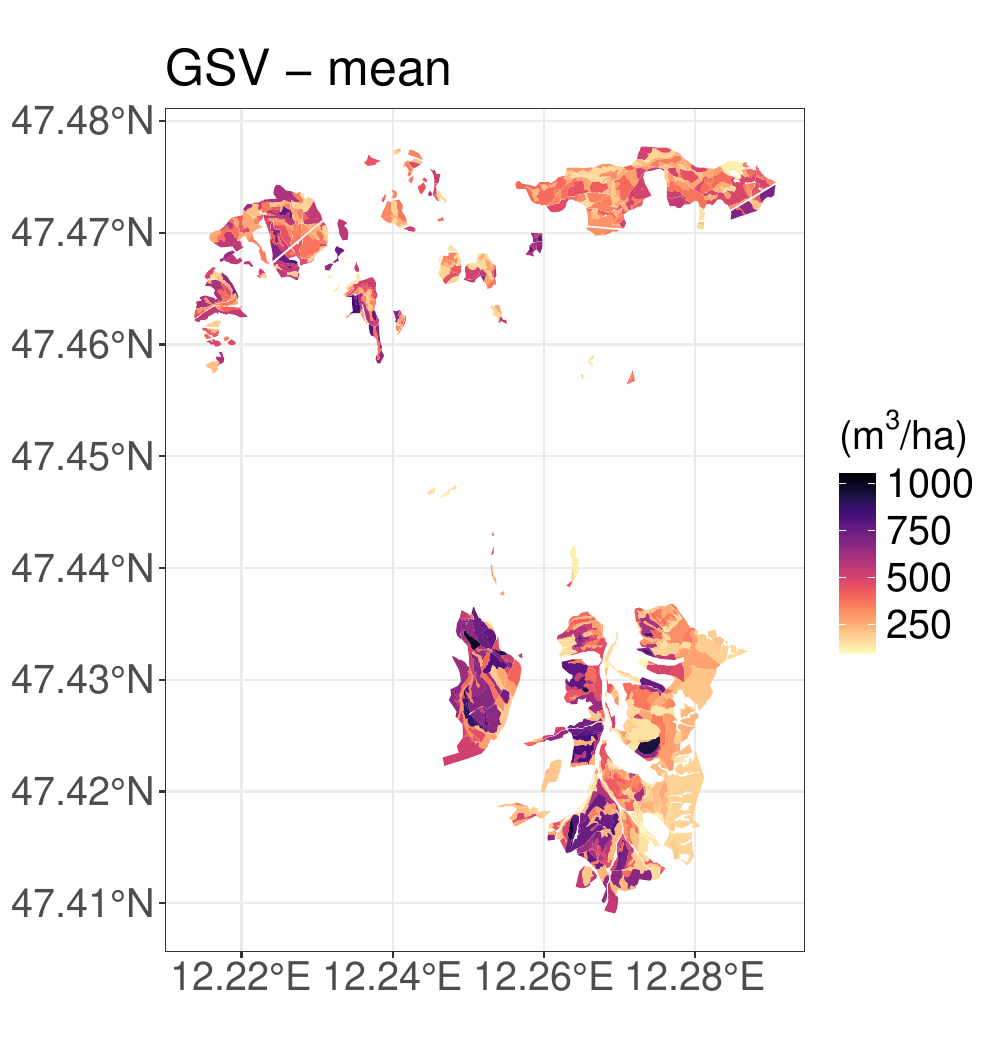}\vspace*{12pt}\\
\includegraphics[width=1\textwidth]{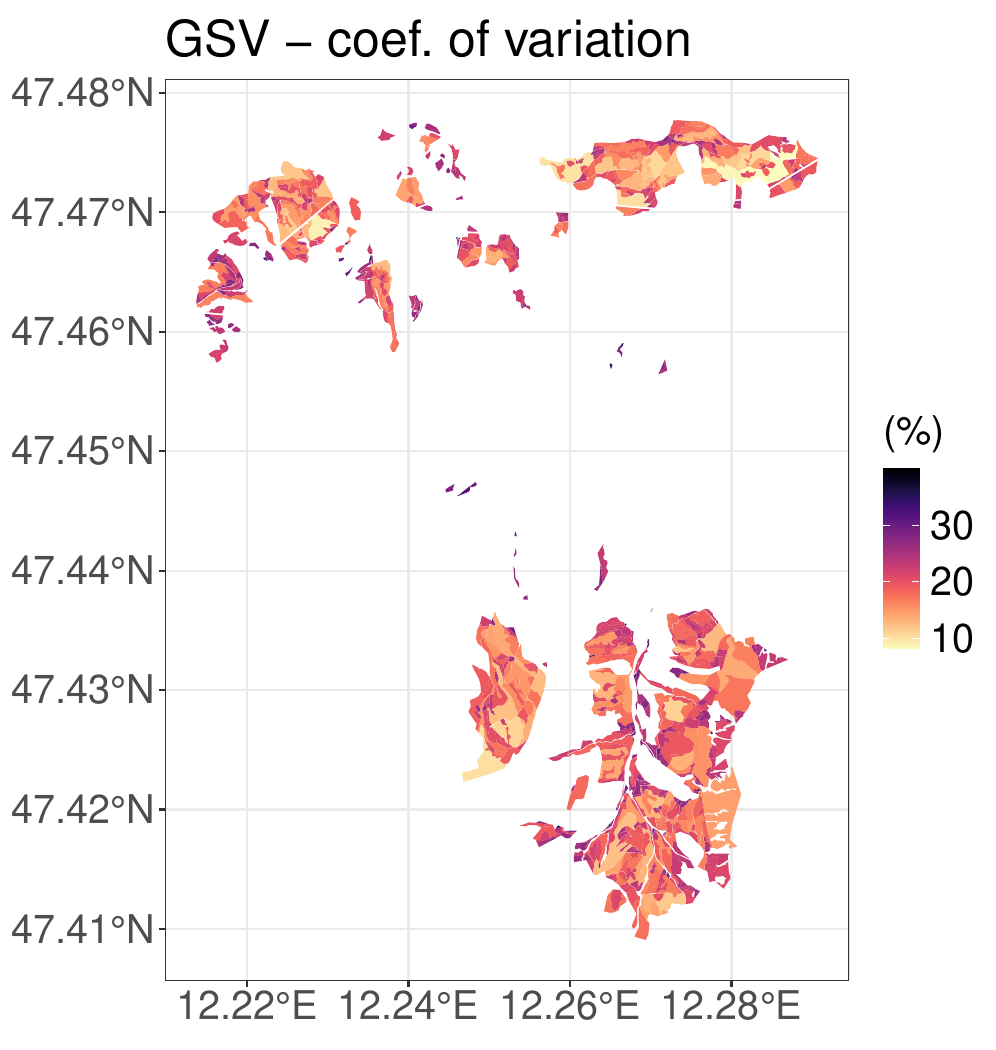}
\end{minipage}\hspace*{10pt}
\begin{minipage}[c]{0.35\textheight}\centering
\includegraphics[width=1\textwidth]{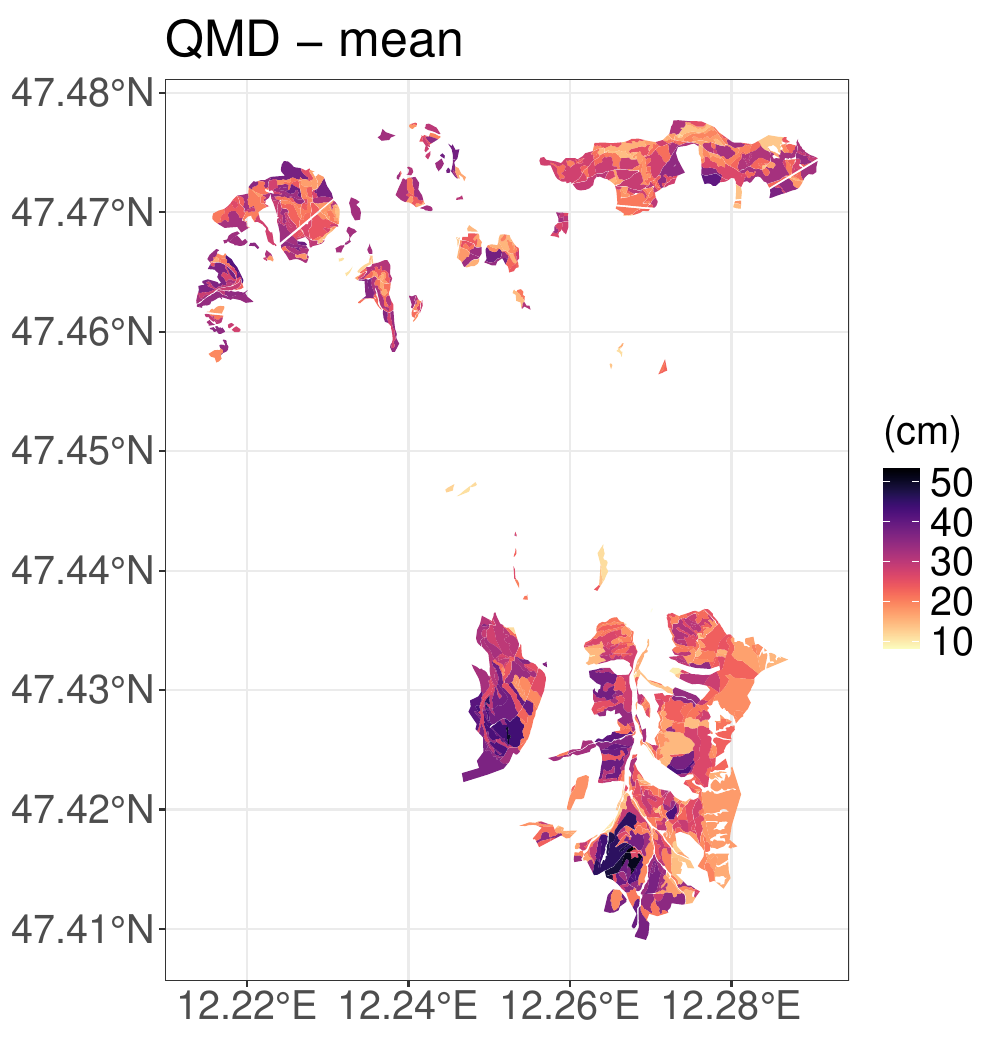}\vspace*{12pt}\\
\includegraphics[width=1\textwidth]{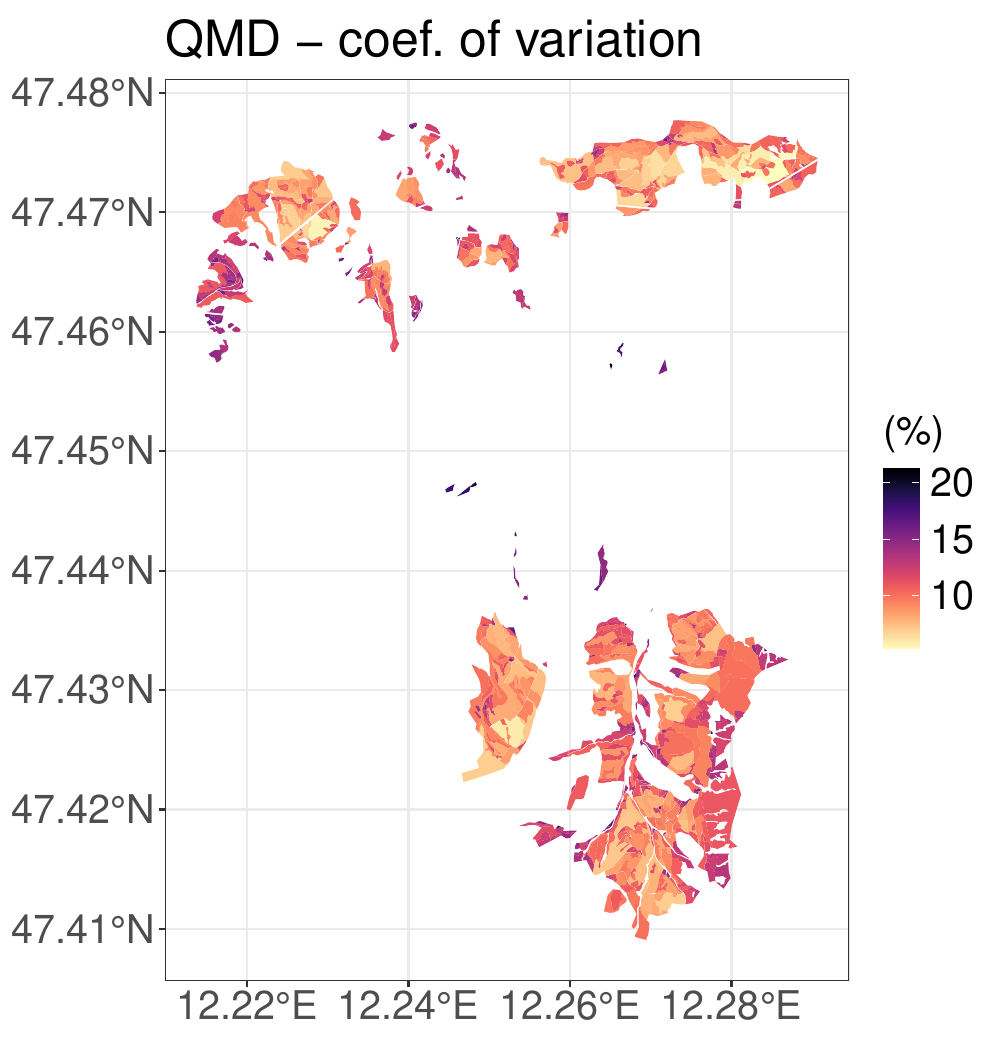}
\end{minipage}\hspace*{10pt}
\begin{minipage}[c]{0.35\textheight}\centering
\includegraphics[width=1\textwidth]{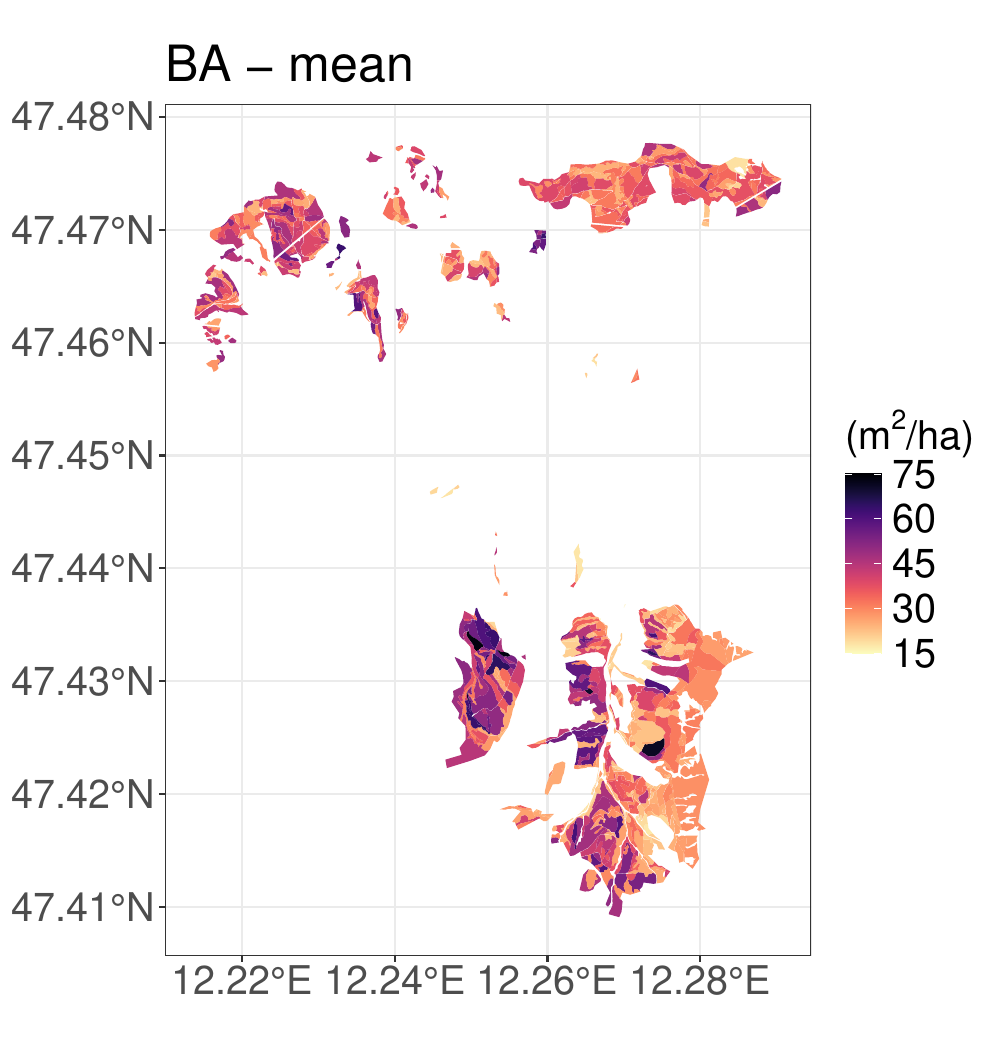}\vspace*{12pt}\\
\includegraphics[width=1\textwidth]{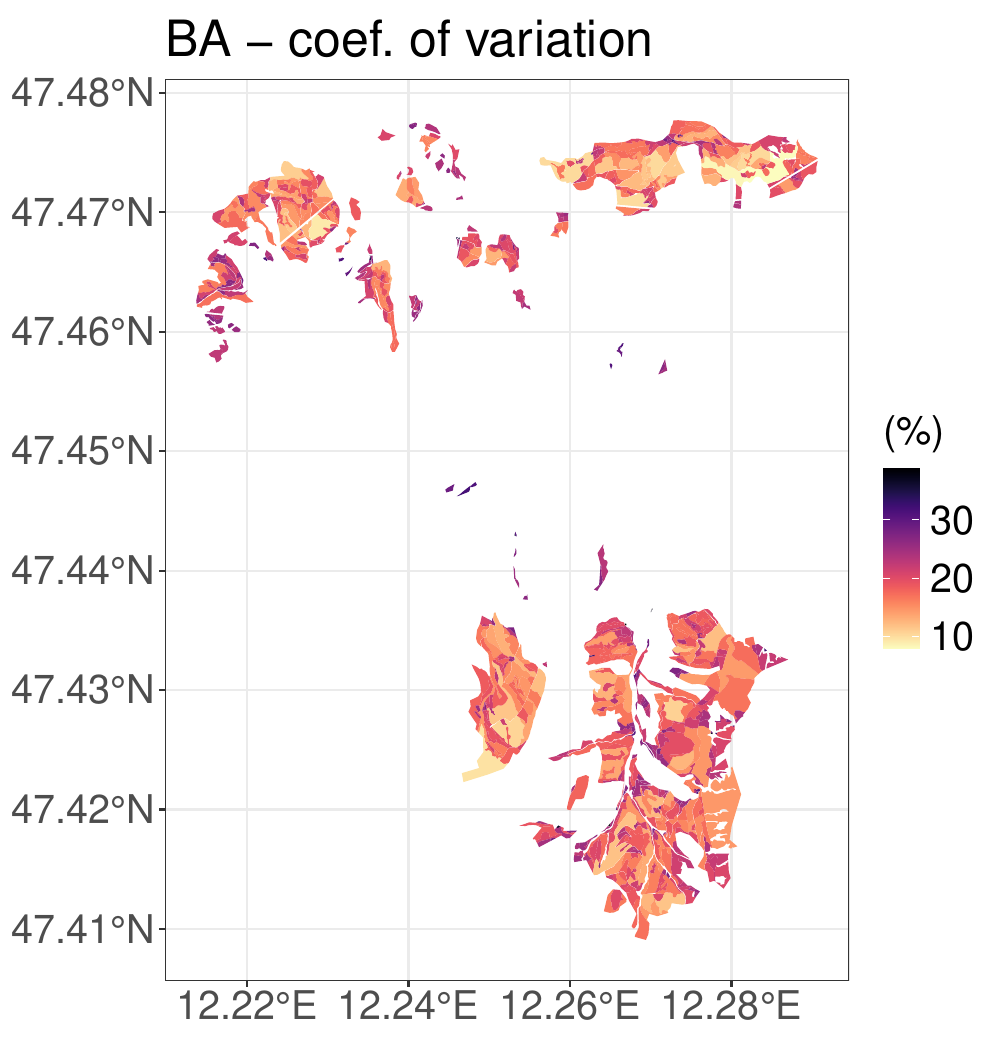}
\end{minipage}\hspace*{10pt}
\begin{minipage}[c]{0.35\textheight}\centering
\includegraphics[width=1\textwidth]{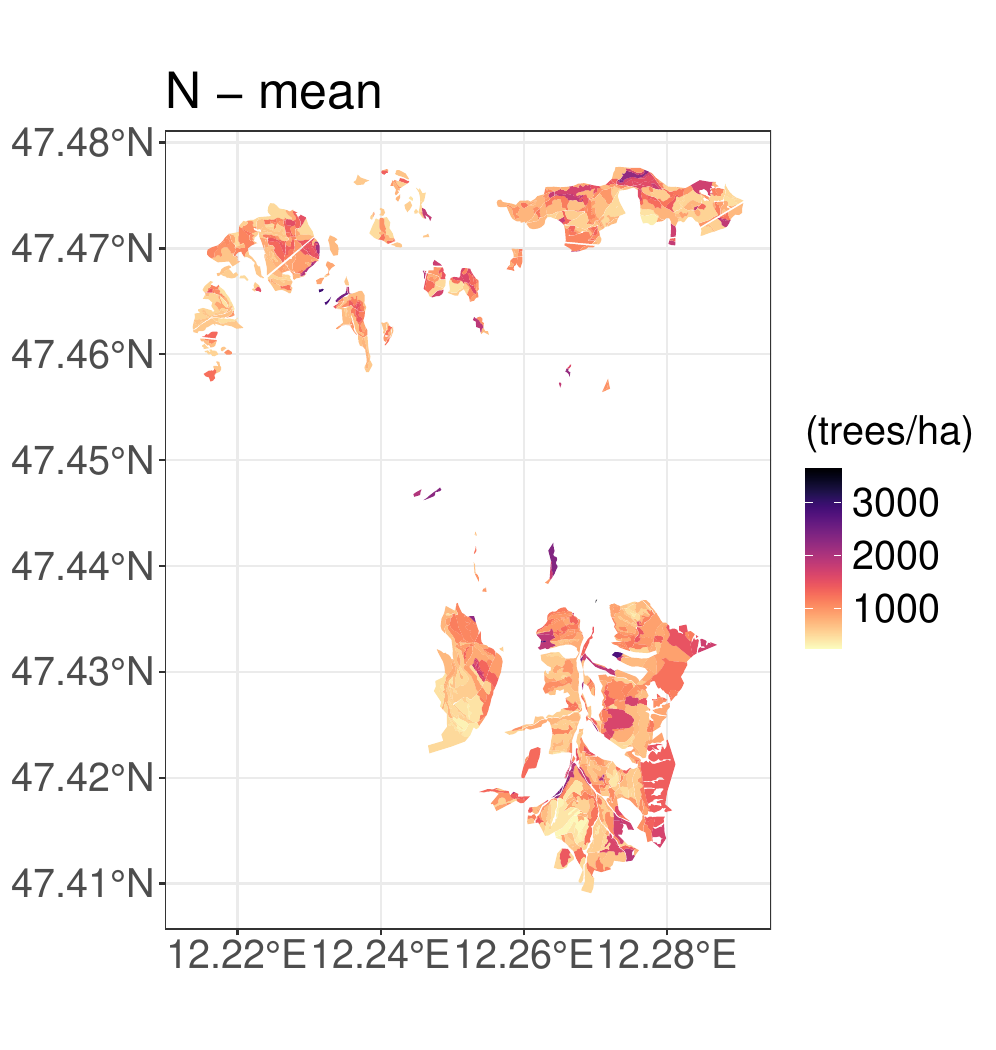}\vspace*{12pt}\\
\includegraphics[width=1\textwidth]{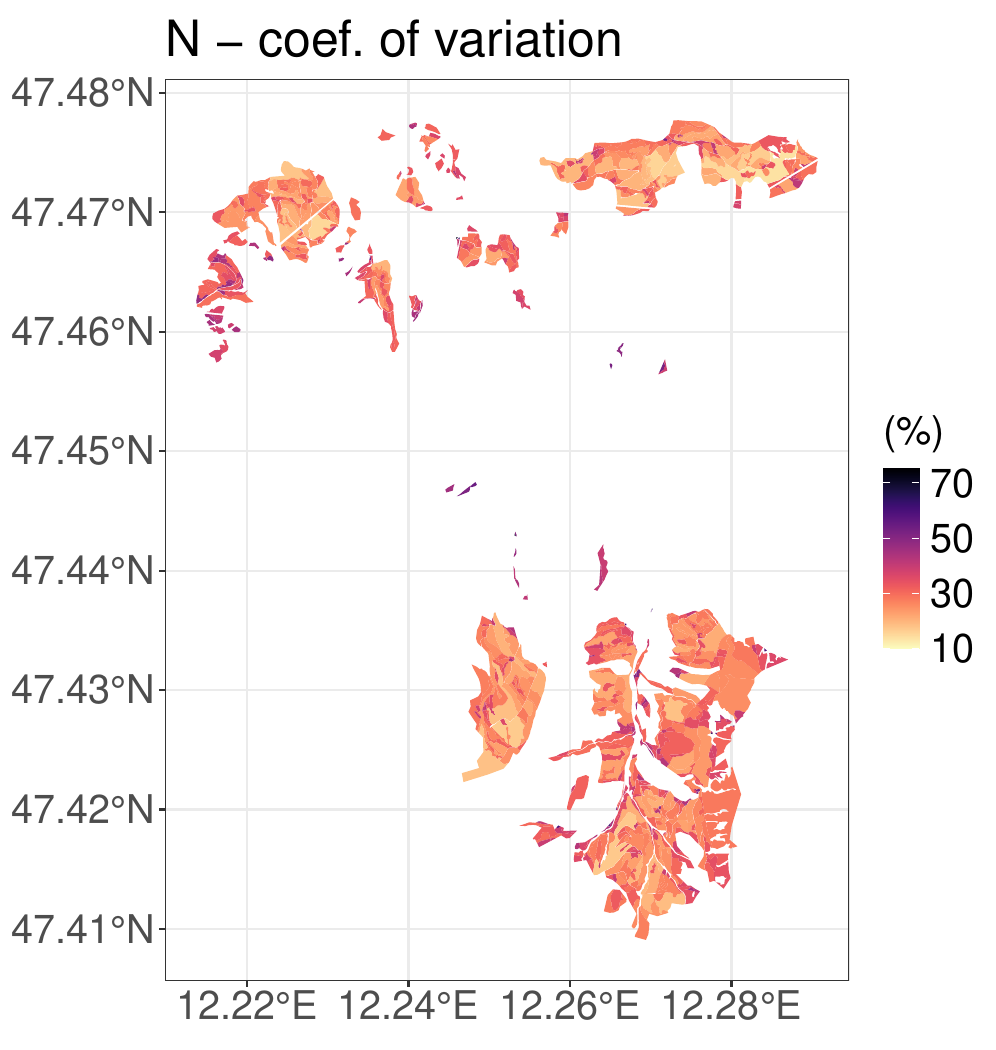}
\end{minipage}
\caption{Posterior predictive distribution (PPD) mean and coefficient of variation for the four outcomes: growing stock timber volume (GSV), quadratic mean stem diameter (QMD), basal area (BA), and stem density (N) for forest stands in Brixen im Thale. Panels in the top row show the PPD mean; panels in the bottom row show the corresponding coefficient of variation.}
\label{fig:spat_pred}
\end{figure*}
\end{landscape}

Stand-level PPD means and CVs are mapped in Figure~\ref{fig:spat_pred}. These maps show relatively high BA and QMD predictions for forest stands located in the southwestern part of the study area. These stands lie in steep terrain with limited forest road access, which restricts timber harvesting. However, due to the siliceous bedrock, these sites are highly productive and associated with high site indices.

In contrast, stands in the southeastern part of the study area—particularly at higher elevations east of the mountain stream ``Brixenthaler Ache''—show lower predicted values for BA, QMD, and GSV. These high-elevation stands have greater stem density and mark a transition to the krummholz zone, dominated by mountain pine (\emph{Pinus mugo}), where only a few monocorm conifers grow. These steeper areas are primarily unproductive protection forests.

In the northern part of the study area, younger and older stands subjected to more intensive silvicultural thinning predominate. This is reflected in lower predicted values for GSV and BA. The terrain here is more moderately sloped and is well served by a dense forest road network due to its proximity to a ski area, enabling more consistent forest management.

The uncertainty maps indicate that the majority of predictions had CVs below 20\% across all four outcomes.

\section{Discussion}\label{sec:discussion}

This study evaluated a suite of univariate and multivariate regression models for SAE in forest inventory, focusing on the estimation of stand-level means for four key forest structure parameters: GSV, BA, QMD, and N. Our results underscore the practical value of model-based SAE approaches, particularly in regions with limited field data and complex terrain.

Across all models, we observed generally low bias and mean percentage bias, indicating that the candidate estimators provide accurate predictions of stand-level means under the cross-validation framework. This is a desirable property for operational use, where unbiasedness is often prioritized alongside precision and interpretability.

We find that spatial random effects modestly improve predictive accuracy when added to models already informed by high-resolution ALS-derived predictors. However, most gains in prediction performance---in terms of RMSPE and interval precision---stem from the inclusion of these predictors rather than the spatial structure. This pattern holds for both univariate and multivariate models.

A key finding is that the spatial dependence observed in our study exhibited relatively short effective ranges, typically around or below 1~km. Much of the spatial structure in the raw measurement data is already accounted for by the ALS-derived predictors, leaving little residual spatial autocorrelation for the random effects to capture. Consequently, the spatial models yield only modest improvements over non-spatial counterparts, as the opportunity for spatial information borrowing is limited. Related studies have reported similar findings regarding the spatial structure of residuals after accounting for remotely sensed predictors. For example, \citet{Mauro2017} evaluated spatial autocorrelation in regression residuals when predicting forest volume, biomass, and QMD using ALS-derived covariates in central Spain. They found that these predictors explained a substantial portion of the spatial structure, leaving only short-range residual autocorrelation. Likewise, \citet{Breidenbach2016} fit spatial regression models to Norwegian National Forest Inventory data using photogrammetrically derived canopy height as a predictor for stand volume. After accounting for stand structure covariates, they observed very short residual spatial ranges---typically less than 25 meters. These findings reinforce the idea that much of the large-scale spatial dependence in forest inventory outcomes can be captured through appropriate predictor variables, leaving only localized residual structure to be modeled stochastically.

The multivariate spatial model performed comparably to its univariate counterparts. This is largely attributable to the fact that all predictions in our evaluation are marginal---each outcome is predicted by integrating over the joint model structure, even in the multivariate case. As a result, there is limited opportunity to exploit cross-outcome correlations during prediction when all outcomes are observed at all locations. 

Although the multivariate models did not yield superior predictions of forest parameters based on standard cross-validation metrics, they did, in some cases, more accurately preserve the correlation structure among outcomes. For example, they were more effective at capturing biologically meaningful relationships, such as the strong correlation between GSV and BA, which reduces the risk of implausible outcome combinations.

Despite the lack of predictive improvement in this fully observed setting, multivariate spatial modeling remains promising for forest inventory applications characterized by spatial misalignment---that is, when only a subset of outcomes is observed at a given location. In such settings, multivariate models provide a principled framework for borrowing strength across outcomes. When supported by sufficient signal in the cross-covariance structure, these models can improve predictions for outcomes that are missing or sparsely sampled (see, e.g., \citealt{Banerjee2014}, Chapter 11; \citealt{Finley2014, Finley2024}).

\section{Conclusion}\label{sec:conclusion}

Our results show the model-based estimators assessed here are particularly relevant to the complex landscape of Brixen im Thale, where terrain, stand structure, and management intensity vary sharply over short distances. ALS-derived predictors provided the primary explanatory power for stand-level attributes, but the random effects captured residual variation attributable to unmeasured site characteristics, management practices, and fine-scale forest structure. While not directly modeled, these latent influences were interpolated across the landscape via the spatial random effects.

The methods presented here support ``precision forestry'' by enabling accurate, uncertainty-quantified, stand-level predictions even in remote and rugged terrain. The approach facilitates efficient forest monitoring and planning, with potential applications in harvest scheduling, ecosystem service valuation, and forest health assessments. Moreover, PPDs generated under the Bayesian framework can be used to propagate uncertainty through optimization routines for operational decision-making, in the same way QMD and BA MCMC samples were used to sample from N's posterior distribution.

For practitioners, the following takeaways are particularly relevant:

\begin{itemize}
  \item ALS-derived predictors are the primary drivers of predictive performance and should be prioritized in model development when available.
  \item Spatial random effects contribute to improved accuracy and uncertainty quantification but offer diminishing returns when predictors already explain most of the spatial structure.
  \item The relatively short spatial correlation scale in this study ($\sim$1~km or less) limits the utility of spatial modeling, especially when strong predictors are present.
  \item Univariate spatial models offer performance comparable to multivariate models when all outcomes are observed, with the added benefit of reduced complexity.
  \item The cross-outcome correlations estimated in the multivariate spatial models can improve estimation of biologically meaningful among outcome correlations.
  \item The Bayesian framework allows flexible transformation and back-transformation of outcomes, enabling modeling on appropriate scales while yielding interpretable predictions.
  \item The Bayesian framework also facilitates sampling from PPDs of derived outcomes such as N given samples from BA and QMD PPDs.
\end{itemize}

Taken together, our findings support the use of model-based SAE methods for generating spatially explicit predictions of forest attributes and their associated uncertainty. Future research will extend spatial regression models by incorporating spatially varying slope coefficients. We also anticipate that the integration of unmanned aerial vehicle (UAV)-based laser scanning with concurrent forest inventory fieldwork will enhance the correlation between outcome and predictor variables. A particular emphasis will be placed on improving the modeling of wood volume at the individual-tree level. Use of PLS data to generate ``digital twins'' of vegetation within sample plots will enable precise 3D measurement of individual tree trunks. This will allow multivariate spatial regression techniques to capture correlations across hierarchical levels, leading to more accurate taper curve predictions along stem axes, improved modeling of within-plot stem shape variability, imputation of missing stem cross-section measurements via posterior predictive simulations, and spatial interpolation across larger study areas. Beyond enhancing estimates of timber volume and growing stock, these developments will facilitate robust aboveground biomass and carbon stock assessments while preserving biologically valid allometric relationships.

\section{Acknowledgments}\label{sec:Acknowledgments}

This study was supported by the project LadiWaldi and was financed by the Austrian Federal Ministry of Agriculture, Forestry, Regions and Water Management within the Waldfonds program under project number 102046. The work of Tobias Ofner-Graff, Valentin Sarkleti, and Philip Svazek was completely financed by LadiWaldi. Finley's work was supported by Michigan State University AgBioResearch and NASA CMS grants Hayes (CMS 2023).

\clearpage

\appendix
\renewcommand{\thefigure}{A.\arabic{figure}}  
\renewcommand{\thetable}{A.\arabic{table}}    
\setcounter{figure}{0}                        
\setcounter{table}{0}                         

\section{Appendix}

\begin{figure}[!h]
\begin{center}
\includegraphics[width=12cm]{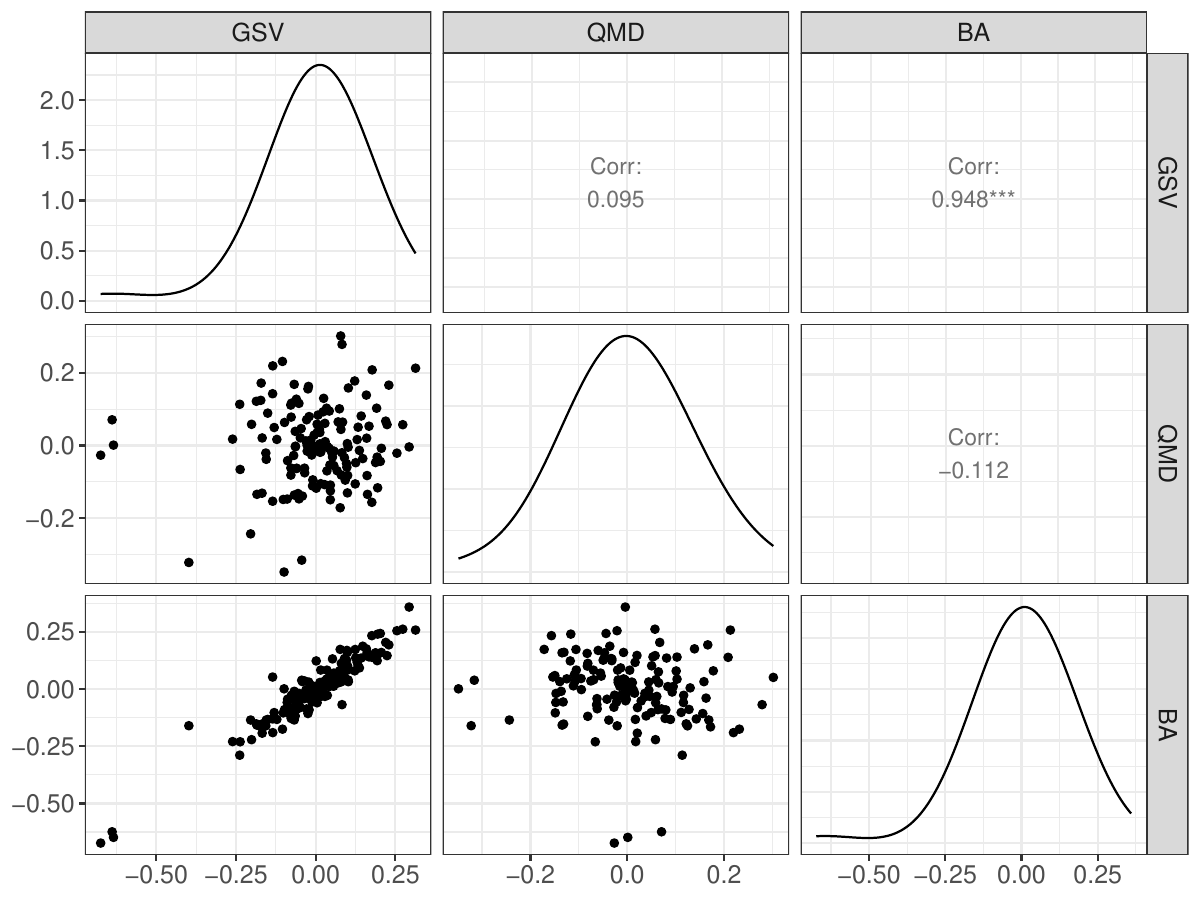}
\caption{Summaries of residuals from univariate spatial all-predictors candidate models. Diagonal panels show the distribution of each model’s residuals; lower triangle panels display pairwise scatter plots between models; and upper triangle panels report the corresponding Pearson correlation coefficients. Asterisks next to the correlation estimates indicate frequentist statistical significance: \textit{**} $p < 0.01$, and \textit{***} $p < 0.001$.}
\label{fig:uni_resid_scatter}
\end{center}
\end{figure}

\begingroup
\setlength{\tabcolsep}{1pt}
\renewcommand{\arraystretch}{.75}
\begin{table}[!h]
\centering
\footnotesize
\begin{tabular}{cccccc}
\toprule
\multicolumn{2}{c}{ } & \multicolumn{2}{c}{Intercept only} & \multicolumn{2}{c}{All predictors} \\
\cmidrule(l{3pt}r{3pt}){3-4} \cmidrule(l{3pt}r{3pt}){5-6}
 & Parameter & Non-spatial & Spatial & Non-spatial & Spatial\\
\midrule
 & $\beta_1$ & \textbf{6.1 (6.0, 6.2)} & \textbf{6.1 (6.0, 6.3)} & \textbf{3.8 (3.1, 4.5)} & \textbf{3.9 (3.1, 4.6)}\\
 & $\beta_{\text{Mean}}$ & - & - & \textbf{0.080 (0.058, 0.10)} & \textbf{0.081 (0.061, 0.10)}\\
 & $\beta_{\text{SD}}$ & - & - & 0.014 (-0.032, 0.059) & 0.023 (-0.023, 0.072)\\
 & $\beta_{\text{P}_{95}}$ & - & - & -0.0014 (-0.030, 0.030) & -0.0052 (-0.038, 0.022)\\
 & $\beta_{\text{Elev.}}$ & - & - & \textbf{0.066 (0.025, 0.11)} & \textbf{0.062 (0.014, 0.11)}\\
\multirow{-6}{*}{\centering\arraybackslash \rotatebox{90}{GSV}} & $\beta_{\text{Asp.}}$ & - & - & -0.019 (-0.11, 0.082) & -0.029 (-0.12, 0.049)\\
\cmidrule{1-6}
 & $\beta_1$ & \textbf{3.3 (3.3, 3.4)} & \textbf{3.4 (3.3, 3.5)} & \textbf{1.6 (1.2, 2.0)} & \textbf{1.6 (1.0, 2.1)}\\
 & $\beta_{\text{Mean}}$ & - & - & \textbf{0.033 (0.020, 0.046)} & \textbf{0.037 (0.023, 0.050)}\\
 & $\beta_{\text{SD}}$ & - & - & \textbf{0.088 (0.061, 0.11)} & \textbf{0.096 (0.069, 0.12)}\\
 & $\beta_{\text{P}_{95}}$ & - & - & -0.012 (-0.028, 0.0065) & -0.016 (-0.034, 0.00079)\\
 & $\beta_{\text{Elev.}}$ & - & - & \textbf{0.056 (0.032, 0.080)} & \textbf{0.057 (0.024, 0.092)}\\
\multirow{-6}{*}{\centering\arraybackslash \rotatebox{90}{QMD}} & $\beta_{\text{Asp.}}$ & - & - & \textbf{0.062 (0.0053, 0.11)} & 0.048 (-0.020, 0.12)\\
\cmidrule{1-6}
 & $\beta_1$ & \textbf{3.7 (3.7, 3.8)} & \textbf{3.7 (3.6, 3.9)} & \textbf{2.2 (1.5, 2.9)} & \textbf{2.1 (1.4, 2.9)}\\
 & $\beta_{\text{Mean}}$ & - & - & \textbf{0.052 (0.030, 0.075)} & \textbf{0.051 (0.031, 0.075)}\\
 & $\beta_{\text{SD}}$ & - & - & -0.033 (-0.081, 0.011) & -0.026 (-0.073, 0.021)\\
 & $\beta_{\text{P}_{95}}$ & - & - & 0.0048 (-0.025, 0.036) & 0.0035 (-0.029, 0.033)\\
 & $\beta_{\text{Elev.}}$ & - & - & \textbf{0.062 (0.020, 0.10)} & \textbf{0.066 (0.017, 0.11)}\\
\multirow{-6}{*}{\centering\arraybackslash \rotatebox{90}{BA}} & $\beta_{\text{Asp.}}$ & - & - & -0.0089 (-0.11, 0.087) & -0.022 (-0.11, 0.067)\\
\bottomrule
\end{tabular}
\caption{Posterior summaries for multivariate candidate models' regression coefficients. Estimates represent the posterior median with 95\% credible interval (CI) bounds in parentheses. Subscripts on regression coefficients indicate the associated predictor variable. Boldface indicates regression coefficients whose credible intervals exclude zero.}\label{tab:multi_beta_params}
\end{table}
\endgroup

\begingroup
\setlength{\tabcolsep}{3.5pt}
\renewcommand{\arraystretch}{.75}
\begin{sidewaystable}[!h]
\footnotesize
\centering
\begin{tabular}{cccccccccc}
\toprule
\multicolumn{2}{c}{ } & \multicolumn{4}{c}{Intercept only} & \multicolumn{4}{c}{All predictors} \\
\cmidrule(l{3pt}r{3pt}){3-6} \cmidrule(l{3pt}r{3pt}){7-10}
\multicolumn{2}{c}{ } & \multicolumn{2}{c}{Univariate} & \multicolumn{2}{c}{Multivariate} & \multicolumn{2}{c}{Univariate} & \multicolumn{2}{c}{Multivariate} \\
\cmidrule(l{3pt}r{3pt}){3-4} \cmidrule(l{3pt}r{3pt}){5-6} \cmidrule(l{3pt}r{3pt}){7-8} \cmidrule(l{3pt}r{3pt}){9-10}
Outcome & Metric & Non-spatial & Spatial & Non-spatial & Spatial & Non-spatial & Spatial & Non-spatial & Spatial\\
\midrule
 & Bias & -6.1 & -7.6 & -7.3 & \textbf{-5.6} & -9.9 & -11. & -9.0 & -8.6\\
 & Bias (\%) & -1.2 & -1.5 & -1.4 & \textbf{-1.1} & -1.9 & -2.1 & -1.8 & -1.7\\
 & RMSPE & 210. & 190. & 210. & 190. & 110. & \textbf{110.} & 110. & \textbf{110.}\\
 & RMSPE (\%) & 41. & 37. & 41. & 38. & 22. & \textbf{21.} & 22. & \textbf{21.}\\
 & 95\% Cover & 96. & 94. & 95. & 92. & 95. & 95. & 95. & 95.\\
\multirow{-6}{*}{\centering\arraybackslash \rotatebox{90}{GSV}} & 95\% CI Range & 1100. & 1000. & 1100. & 1000. & 670. & 670. & 670. & \textbf{650.}\\
\cmidrule{1-10}
 & Bias & \textbf{-0.049} & -0.13 & -0.093 & -0.081 & -0.079 & -0.20 & -0.089 & -0.15\\
 & Bias (\%) & \textbf{-0.17} & -0.43 & -0.31 & -0.27 & -0.27 & -0.66 & -0.30 & -0.49\\
 & RMSPE & 7.2 & 6.9 & 7.2 & 6.9 & 4.2 & \textbf{4.1} & 4.3 & 4.2\\
 & RMSPE (\%) & 24. & 23. & 24. & 23. & 14. & \textbf{14.} & 14. & 14.\\
 & 95\% Cover & 95. & 95. & 94. & 95. & 95. & 95. & 96. & 93.\\
\multirow{-6}{*}{\centering\arraybackslash \rotatebox{90}{QMD}} & 95\% CI Range & 37. & 36. & 37. & 35. & 22. & 23. & \textbf{22.} & \textbf{22.}\\
\cmidrule{1-10}
 & Bias & -0.35 & -0.36 & -0.48 & \textbf{-0.34} & -0.64 & -0.81 & -0.64 & -0.63\\
 & Bias (\%) & -0.78 & -0.80 & -1.1 & -0.77 & -1.4 & -1.8 & -1.4 & -1.4\\
 & RMSPE & 14. & 13. & 14. & 13. & 9.9 & \textbf{9.4} & 9.9 & \textbf{9.4}\\
 & RMSPE (\%) & 32. & 30. & 32. & 30. & 22. & \textbf{21.} & 22. & \textbf{21.}\\
 & 95\% Cover & 95. & 94. & 95. & 95. & 97. & 97. & 97. & 95.\\
\multirow{-6}{*}{\centering\arraybackslash \rotatebox{90}{BA}} & 95\% CI Range & 77. & 72. & 77. & 72. & 57. & 57. & 57. & \textbf{56.}\\
\cmidrule{1-10}
 & Bias & -71. & -68. & \textbf{-21.} & -28. & -27. & -33. & -28. & -37.\\
 & Bias (\%) & -9.0 & -8.6 & \textbf{-2.6} & -3.5 & -3.4 & -4.2 & -3.6 & -4.7\\
 & RMSPE & 390. & 400. & 380. & 380. & \textbf{310.} & \textbf{310.} & \textbf{310.} & \textbf{310.}\\
 & RMSPE (\%) & 50. & 50. & 49. & 48. & \textbf{39.} & \textbf{39.} & \textbf{39.} & \textbf{39.}\\
 & 95\% Cover & 98. & 99. & 96. & 95. & 93. & 93. & 93. & 92.\\
\multirow{-6}{*}{\centering\arraybackslash \rotatebox{90}{N}} & 95\% CI Range & 2700. & 2600. & 2200. & 2200. & \textbf{1600.} & \textbf{1600.} & 1700. & 1700.\\
\bottomrule
\end{tabular}
\caption{Unit-level cross-validation summary for candidate models. Where applicable, the best-performing value for each metric is shown in bold. In cases where multiple models yield similar performance, more than one value may be bolded.}\label{tab:kfold_unit_metrics}
\end{sidewaystable}
\endgroup

\begin{table}[htbp]
\centering
\caption{Correlation estimates between outcomes based on the cross-validation for intercept-only models.}\label{tab:cor_int}
\begin{subtable}[t]{0.45\textwidth}
  \centering
  \caption{Univariate non-spatial}
  \label{tab:cor_int_a}
  \begin{tabular}{lllll}
  \toprule
  & \text{GSV} & \text{QMD} & \text{BA} & \text{N} \\
\text{GSV} & 1.0 &  &  & \\
\text{QMD} & 0.0030 & 1.0&  & \\
\text{BA} & -0.0017 & 0.0012 & 1.0 & \\
\text{N} & -0.0050 & -0.65 & 0.51 & 1.0 \\
  \bottomrule
  \end{tabular}
\end{subtable}
\hfill
\begin{subtable}[t]{0.45\textwidth}
  \centering
  \caption{Univariate spatial}
  \label{tab:cor_int_b}
  \begin{tabular}{lllll}
    \toprule
  & \text{GSV} & \text{QMD} & \text{BA} & \text{N} \\
\text{GSV} & 1.0 &  &  & \\
\text{QMD} & 0.043 & 1.0 &  & \\
\text{BA} & 0.090 & 0.022 & 1.0 & \\
\text{N} & 0.0091 & -0.65 & 0.49 & 1.0\\
  \bottomrule
  \end{tabular}
\end{subtable}
\vskip1em
\begin{subtable}[t]{0.45\textwidth}
  \centering
  \caption{Multivariate non-spatial}
  \label{tab:cor_int_c}
  \begin{tabular}{lllll}
  \toprule
  & \text{GSV} & \text{QMD} & \text{BA} & \text{N} \\
\text{GSV} & 1.0&  &  & \\
\text{QMD} & 0.46$^\ast$ & 1.0&  & \\
\text{BA} & 0.93$^\ast$ & 0.22$^\ast$ & 1.0& \\
\text{N} & 0.12$^\ast$ & -0.64 & 0.38$^\ast$ & 1.0\\
  \bottomrule
  \end{tabular}
\end{subtable}
\hfill
\begin{subtable}[t]{0.45\textwidth}
  \centering
  \caption{Multivariate spatial}
  \label{tab:cor_int_d}
  \begin{tabular}{lllll}
  \toprule
  & \text{GSV} & \text{QMD} & \text{BA} & \text{N} \\
\text{GSV} & 1.0 &  &  & \\
\text{QMD} & 0.45$^\ast$ & 1.0 &  & \\
\text{BA} & 0.92$^\ast$ & 0.20$^\ast$ & 1.0 & \\
\text{N} & 0.12$^\ast$ & -0.65 & 0.39 & 1.0\\
  \bottomrule
  \end{tabular}
\end{subtable}
\caption*{\small Values are posterior means. An asterisk (\,$^\ast$) indicates that the 95\% credible interval includes the corresponding empirical correlation given in Table~\ref{tab:sampleplot_cor}.}
\end{table}

\clearpage
\bibliographystyle{apalike}
\bibliography{multispatreg}

\end{document}